\documentclass{elsart}

\usepackage{icarus}

\usepackage{natbib}

\bibpunct{(}{)}{;}{a}{,}{,}

\usepackage{graphicx}

\usepackage{amssymb}

\usepackage{amsmath}

\def\comm#1           {{\tt (COMMENT: #1)}}

\begin{document}

\begin{frontmatter}


\title{Ensemble Properties of Comets in the Sloan Digital Sky Survey}

\author[Adler]{Michael Solontoi}, 
\author[UW]{\v{Z}eljko Ivezi\'c},
\author[hubble,harvard]{Mario Juri\'{c}}
\author[UW]{Andrew C. Becker},
\author[UW]{Lynne Jones},
\author[West]{Andrew A. West},
\author[Kent]{Steve Kent}
\author[Prin]{Robert H. Lupton},
\author[UW]{Mark Claire},
\author[Prin]{Gillian R. Knapp},
\author[UW]{Tom Quinn},
\author[Prin]{James E. Gunn},
\author[PennS]{Donald P. Schneider}

\address[Adler]{Adler Planetarium, 1300 S. Lake Shore Drive, Chicago, IL 60605, USA}
\address[UW]{University of Washington, Dept. of Astronomy, Box 351580,
  Seattle, WA 98195, USA}
\address[hubble]{Hubble Fellow}
\address[harvard]{Harvard College Observatory, Cambridge, MA 02138, USA}
\address[Prin]{Princeton University Observatory, Princeton, NJ 08544, USA}
\address[West]{Department of Astronomy, Boston University, 725 Commonwealth Ave, Boston, MA 02215, USA}
\address[Kent]{Fermi National Accelerator Laboratory, Batavia, IL
  60510, USA}
\address[PennS]{Department of Astronomy and Astrophysics,
  Pennsylvania State University, University Park, PA 16802, USA}



\end{frontmatter}

\begin{flushleft}
\vspace{1cm}
Number of pages: \pageref{lastpage} \\
Number of tables: \ref{lasttable}\\
Number of figures: \ref{lastfig}\\
\end{flushleft}

\begin{pagetwo}{Ensemble Properties of Comets in the Sloan Digital Sky Survey}

Michael Solontoi \\
Adler Planetarium\\
1300 S. Lake Shore Drive, Chicago, IL 60605\\
\\
Email: msolontoi@adlerplanetarium.org\\

\end{pagetwo}

\begin{abstract}

We present the ensemble properties of 31 comets (27
resolved and 4 unresolved) observed by the Sloan Digital Sky Survey
(SDSS).  This sample of comets 
    represents about 1 comet per 10 million SDSS photometric objects.
    Five-band ($u,g,r,i,z$) photometry is used to determine the comets' colors,
    sizes, surface brightness profiles, and rates of dust production in
    terms of the A$f\rho$ formalism.  We find that the cumulative
    luminosity function for the Jupiter Family Comets in our sample is well
    fit by a power law of the form $N(<H) \propto 10^{(0.49
    \pm 0.05) H}$ for $H<18$, with evidence of a much shallower fit $N(<H) \propto 10^{(0.19
    \pm 0.03) H}$ for the faint ($14.5<H<18$) comets. The resolved comets show an
    extremely narrow distribution of colors ($0.57 \pm 0.05$ in $g-r$ for
    example), which are statistically indistinguishable from that of
    the Jupiter Trojans.  Further, there is no
    evidence of correlation between color and physical, dynamical, or
    observational parameters for the observed comets.
\end{abstract}


\section {Introduction}

The physical properties of small planetary bodies offer
insight into the formation and evolution of the Solar System.  However, \emph{in situ}
observational studies of the remote regions of the Solar System are limited to
the largest bodies in the Kuiper Belt, and are still not possible in the case of
the Oort Cloud.  Nevertheless, through delivery of
scattered members of their populations, such as comets, these regions may be
probed observationally.  Comets may be
distinguished from other populations of small bodies due to their
activity: the production of a gas and dust comae, typically when the comet is at small
heliocentric distances, due to the sublimation of volatiles.  The
comet populations that are thought to directly sample these two remote
regions are the Jupiter Family Comets (JFC), from the Kuiper Belt,
characterized by generally low-inclination prograde orbits, and Long
Period Comets (LPC) thought to originate in the Oort Cloud, with a
large range of orbital inclinations, directions, and whose aphelia
lie far beyond the orbits of the planets \citep{2004come.book..659J}.  Interactions with the giant
planets can give rise to comets that exist in-between these two
populations, comets with orbits like that of 1P/Halley. Dynamically,
the Tisserand parameter with respect
to Jupiter is often employed to distinguish between the various
populations of comets based on their orbits
\citep{1997Icar..127...13L}. Recently a
new population of comets has been discovered within the main asteroid
belt \citep{2006Sci...312..561H};  the orbits of these objects are
indistinguishable from those of main belt asteroids, but they have
observed activity in the form of comae and dust trails.

Building upon the methodology of detecting active comets in the SDSS
described in our previous work \citep{2010Icar..205..605S}, we
present here the analysis of properties derived from 35 sets of $u,g,r,i,z$ band photometry of
31 comets observed by the Sloan Digital Sky Survey
\citep{2000AJ....120.1579Y}.  This 
includes comets found through the methodology described by \cite{2010Icar..205..605S},  
along with those recovered through orbital algorithms
developed for matching asteroids found in the SDSS to predicted
positions of known asteroids \citep{2002SPIE.4836...98I, 2002AJ....124.1776J}.

In Section \ref{method} we present a brief overview of the methods employed to
find comets in the SDSS.  The data
and analysis of the observations of both the resolved comets (showing
comae), and those that were unresolved (point-source like) are
discussed in Section \ref{analysis}.  We discuss the 
color, size and dust properties of the individual comets, as
well as the cumulative luminosity and size distributions of the
Jupiter Family Comets, in Section \ref{disc}.

\section{Observations}\label{method}

\subsection{Brief Overview of the SDSS}

The data presented here are based on the Seventh SDSS Public Data
Release, hereafter DR7 \citep{2009ApJS..182..543A}, which ran through
July 2008, and contains over 357 million unique photometric objects.
Detailed information about
this data release including sky coverage, changes from previous data
releases and data quality statistics can be found at
http://www.sdss.org/DR7, and in \cite{2009ApJS..182..543A}. Of
particular interest to Solar System studies, the 
survey covers the sky at and near the ecliptic from approximately
$\lambda = 100^{\circ}$ to $\lambda = 225^{\circ}$.  The repeat
scans of the Southern Galactic hemisphere (Stripe 82; crossing $\lambda =
0^{\circ}$) also pass through the Ecliptic.

The SDSS is a digital photometric and spectroscopic survey that
covered about one quarter of the Celestial Sphere in the North Galactic
cap and a smaller ($\sim 300$ deg$^2$) but much deeper survey
in the Southern Galactic hemisphere and began standard operations in
April 2000
\citep[see][]{2000AJ....120.1579Y, 2002AJ....123..485S, 2003AJ....126.2081A, 2004AJ....128..502A,
  2005AJ....129.1755A, 2009ApJS..182..543A, 2006ApJS..162...38A,
  2008ApJS..175..297A}.  SDSS used a dedicated 2.5m 
  telescope \citep{2006AJ....131.2332G} to
  provide homogeneous and deep (r $<$ 22.5) photometry in five
  band passes \citep{1996AJ....111.1748F, 1998AJ....116.3040G,
  2001AJ....122.2129H, 2002AJ....123.2121S, 2006AN....327..821T}
  repeatable to 0.02 mag 
  \citep[root-mean-square scatter, hereafter rms, for sources not limited by
  photon statistics,][]{2003MmSAI..74..978I} and with a zero-point
  uncertainty of $\sim$0.02-0.03 mag \citep{2004AN....325..583I}. The
  flux densities of detected objects were measured almost simultaneously
  in five bands (\emph{u, g, r, i}, and \emph{z}) with effective
  wavelengths of 3540 \r{A}, 4760 \r{A}, 6280 \r{A}, 7690 \r{A}, and
  9250 \r{A} \citep{2010AJ....139.1628D}.  The large sky coverage of the survey (almost 12,000
  deg$^{2}$ of sky) has resulted in photometric measurements of
  approximately 357 million objects \citep{2009ApJS..182..543A}.
  The completeness of SDSS catalogs for
  point sources is $\sim$99.3\% at the bright end and drops to 95\% at
  magnitudes of 22.1, 22.4, 22.1, 21.2, and 20.3 in \emph{u, g, r, i},
  and \emph{z}, respectively.  Astrometric positions are accurate to
  better than 0.1 arc-second per coordinate (rms) for sources with r
  $<$ 20.5 \citep{2003AJ....125.1559P}, and the morphological
  information from images allows reliable star-galaxy separation to r
  $\sim$ 21.5 \citep{2002SPIE.4836..350L, 2002ApJ...579...48S}. A compendium
  of the technical details about SDSS can be found on the SDSS web
  site (http://www.sdss.org), which also provides the interface for
  public data access.

\subsection{SDSS Photometric Measurements}
When discussing active (resolved) comets, the magnitudes used
here are the ``model magnitudes" measured by the SDSS. These are
designed for galaxy photometry, and are determined by
accepting the better of a de Vaucouleurs and an exponential profile of
arbitrary size and orientation \citep[see][]{2002AJ....123..485S}.
While the surface brightness profile of a comet differs
from that of a galaxy, \cite{2010Icar..205..605S} demonstrate that these
models still produce good
fits to the cometary data.  For the unresolved comets, the magnitudes are based
on their point spread function (PSF) magnitude which is
measured by fitting the point
spread function model to the object.  These two magnitudes serve as
the basis for discriminating between resolved (galaxy-like) and
unresolved (star-like) sources in the SDSS; if the difference in 
magnitude between the $r$ band PSF magnitude and the $r$ band model
magnitude is
greater than 0.145 then SDSS classifies the object as resolved, or as
a ``galaxy'' type object \citep{2002AJ....123..485S}.

\subsection{Finding Moving Objects in the SDSS Data}

Although mainly designed for observations of extragalactic
sources, the SDSS is significantly contributing to studies of the solar system,
notably in the success it has had with asteroid detections, cataloged
in the SDSS Moving Object
Catalog \citep[hereafter SDSS MOC,][]{2001AJ....122.2749I}.  This
public, value-added, catalog of
SDSS asteroid observations contains, as of its fourth release,
471,000 measurements of moving objects, 220,000 of which have been matched
to 104,000 known asteroids from the ASTORB file\footnote{see
  ftp://ftp.lowell.edu/pub/elgb/astorb.html.}.  The SDSS MOC data is
of high quality, and has been widely used in recent studies of
asteroids \citep[see][]{2001AJ....122.2749I, 2002AJ....124.1776J,
  2007LPI....38.1851B, 2008Icar..198..138P, 2008A&A...488..339A,
  2010A&A...510A..43C}. 

The SDSS camera \citep{1998AJ....116.3040G} uses a drift-scan-like
technique along great circles and detects objects in the
order \emph{r, i, u, z, g}, with detection in two successive bands separated in
time by 72 seconds, with the different
bands registered to the same coordinate system using stationary
stars \citep{2003AJ....125.1559P}. Moving objects appear to have their
photocenters spacially separated in different filters when color composite images are made. An
angular velocity, calculated from the astrometrically-calibrated image
centroids in each filter, is calculated for every photometric object. 

One of the features that allows the automatic creation of the SDSS
MOC is that asteroids appear as point sources.  A similar approach to the SDSS
observations of resolved objects, in order to find comets, results in a
sample well in excess of a million candidates
\citep{2010Icar..205..605S}, vastly larger than even the most optimistic estimate of observable comets.  Even if every known
comet in the sky were to be imaged, this sample would still be
dominated by false
positives by a factor of about 1000:1.  Therefore, extended objects
pose a much more difficult challenge for an automated
reduction pipeline, further complicated by the fact
that we are searching for \emph{moving} extended objects, which to the SDSS
photometric pipeline are interpreted as ``moving galaxies.''

\subsection{Finding Comets in the SDSS Data}
In order to successfully acquire comet observations from the SDSS
database we employed two methods.  The first involved making
selection cuts based on SDSS measured photometry and data processing
quality flags.  The resulting candidate objects were then visually
inspected (this method is discussed in depth by
\citealt{2010Icar..205..605S}).  The advantage
of this technique is that it is blind to the known comet sample.  It does
not matter whether the candidates are known comets,
thought to be asteroids, or being observed for the first
time (i.e. being discovered).  The weakness of this approach is that it requires visual identification of
each candidate object, and is subject to SDSS pipeline issues (e.g. deblending
errors, multiple epochs represented by a single image; see \citealt{2010Icar..205..605S} for
details). Since this technique was specifically designed for
``cometary'' (resolved) objects it will not select comets that are
inactive, or have sufficiently low activity to appear as point
sources at the resolution of the SDSS. 

A second technique to select comets from the SDSS employs the method
used by \cite{2002AJ....124.1776J} to identify 
known asteroids in the SDSS MOC.  We
utilized the code developed by \cite{2002AJ....124.1776J} to propagate
the orbits of known comets through the SDSS
observational cadence. Much as in the case of the SDSS MOC, this code generates all
possible RA and dec locations that a known comet could have been observed by the
telescope over the course of the survey.  This method enabled the
selection of both resolved and unresolved comets. Visual
inspection of candidates is still required
as this list of positions sometimes has large astrometric uncertainties.
Comets, unlike asteroids, have significant
non-gravitational effects on their motion due to out-gassing
during perihelion passage that alter the orbit
of a comet over a given apparition, and certainly from passage to
passage.  For example, Comet Encke returns on average about 2.5 hours
sooner than predicted by gravitational computations, a phenomenon that
was noticed as far back as the 19th
century \citep{1950ApJ...111..375W}, and subsequent studies have
demonstrated that the majority of periodic comets experience
non-gravitational acceleration
\citep{1968BAICz..19..351S}.  This fact, coupled with many comets not having
well calibrated observations over several orbits, leads to positional
uncertainties in the predicted position.  The comet can be too far from the predicted
position to be uniquely matched by the algorithm, and
in some cases it may not even be present in the SDSS observation field.  A
second weakness is that the underlying orbital code, based on OrbFit
\citep{1999Icar..137..269M} can only be used to find comets with bound
(elliptical) orbits, and thus excludes Long Period Comets.  The final (and obvious) 
limitation is that it will only select known comets. 

Utilizing these two methods; 35 observations of 31
comets have been identified in the SDSS dataset, including two comets
discovered by
the survey (C/1999 F2 Dalcanton and C/2000 QJ46 LINEAR). Two additional new
comets were observed, but have not been successfully matched to any
solar system body with a determined orbit, and thus will not be discussed
in this paper; photometry of those two objects is described by
\cite{2010Icar..205..605S}. 

 While 35 observations of 31 comets seems small in number compared the typical large number of objects associated with the SDSS, they are in line with what is expected.  As discussed in \cite{2010Icar..205..605S}, SDSS should be thought of as a single epoch survey in terms of comet results.  While different comets are observed on different nights, the sky coverage for a single night is fairly small compared the entire sky, and the cadence of scans is not optimized to follow individual solar system objects.  The usable area of the survey for comet work is about 20\% of the whole sky.  How many comets should we expect to find in such a survey?  The Minor Planet Center\footnote{http://www.minorplanetcenter.org/iau/mpc.html} lists approximately 200 comets bright enough for robust detection ($r<20$) on any given night. Taking these 200 potentially observable comets as a typical snapshot of the sky there should be approximately 40 comets in the survey (200 comets on the sky * 20\% of the sky).  Using SDSS DR5 to estimate our completeness \citep{2010Icar..205..605S} we found it to be about 80\%.  Taking the sky coverage of SDSS and out completeness into account we expect to find approximately 32 comets, a prediction consistent with 35 observations of 31 comets reported here

Table \ref{table_obs} lists the SDSS $r$
band magnitudes, the difference between the measured PSF and model magnitudes (a point source would be equal to 0) and observing geometry  (geocentric and heliocentric distances, and phase angle) for each of the comets discussed in this work.

\section {Analysis}\label{analysis}

Here we discuss the methods for determining the photometric colors and
surface brightness profiles for the
resolved comets.  From these measurements, estimates can be made on
the upper limit for nuclear radius, and of the dust production via the
quantity A$f\rho$ \citep{1984AJ.....89..579A}.   We
then discuss the measurement of photometric colors and radii for the
unresolved  comets in section \ref{s_unresolved}.

\subsection {Resolved Comets}\label{s_resolved}
In this sample, 31 observations of 27 comets are classified by the SDSS
as resolved sources, having a $r$ band PSF$-$model magnitude value greater
than 0.145.
For each of these comets the corrected frames (fits files) were
obtained from the SDSS Data Archive
Server\footnote{http://das.sdss.org/www/html/} (DAS) for all five filters.  Nearby stars
and galaxies were masked and
radial surface brightness profiles were extracted independently in
all five bands for each comet using routines in the
IDP3\footnote{http://mips.as.arizona.edu/MIPS/IDP3} image manipulation
package. These profiles consist of azimuthally averaged concentric
annuli about the optocenter of the comet in each photometric band.
This allows four colors ($u-g$, $g-r$, $r-i$, and  $i-z$) to
be determined for all comets, even those whose SDSS assigned magnitudes are
questionable due to photometric processing errors and non-optimal intensity profile for the flux measurement  \cite[see][sections 2.3, 3.4 and references therein]{2010Icar..205..605S}.  In the case of comets with well measured photometry, the total integrated magnitudes determined matched those reported by the SDSS photometric pipeline.

As common with broad band filter studies of comets, we interpret 
the measured flux and resulting colors as due to the dust coma.  Active comets are not free of gas however, and potentially strong emission lines may be present particularly in the wavelength ranges of the $u$ and $g$ bands.  To verify these gas emission lines make a negligible contribution to the integrated flux (and therefore color) a sample comet spectrum \citep{2006DPS....38.1201H} was convolved with the SDSS filter transmission curves \citep{2010AJ....139.1628D} to produce $u$, $g$, and $r-$band magnitudes.  Compared to a ``smoothed'' version of the spectrum (one with the emission lines removed) individual magnitudes changed by $3\%$, $0.1\%$, and $<0.1\%$ for $u$, $g$, and $r-$band magnitudes respectively.  These changes  less than the standard deviation of the measured model magnitudes for the comets in each of these filters.

\subsubsection{Photometric Colors of Resolved Comets}
\begin{figure} [p]
\begin{center}
\includegraphics[width=.5\textwidth]{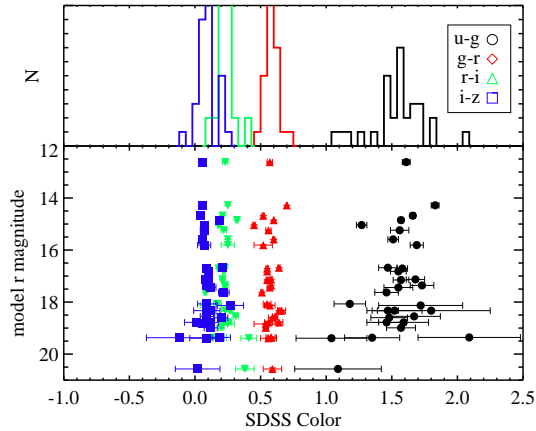} 
\end{center}
\caption[Color-magnitude diagram for the resolved comets]{Color-magnitude distribution of the resolved comets with
  2-$\sigma$ error bars.  The
plotted symbols are: $\bigcirc$ $u-g$ (black), $\triangle$ $g-r$ (red),
  $\triangledown$
$r-i$ (green), and $\square$ $i-z$ (blue).  Plotted above is the
  histogram of the color distribution.}
\label{fig_color_hist} 
\end{figure}

\begin{figure} [p]
\begin{center}
\includegraphics[width=0.75\textwidth]{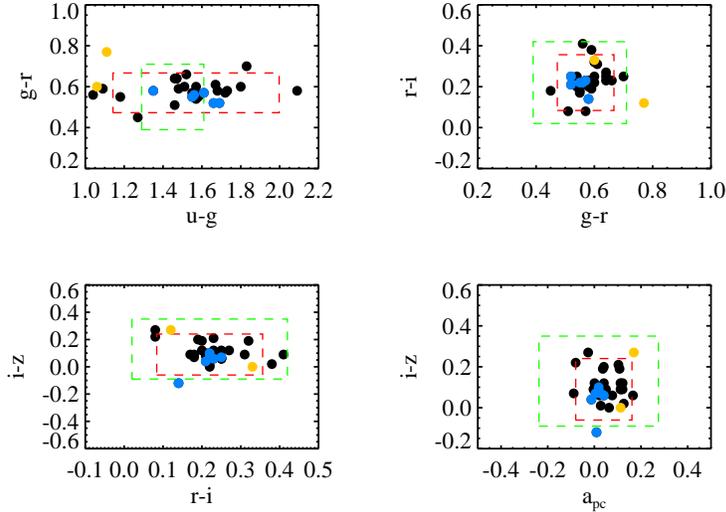} 
\end{center}
\caption[Color-Color plots version 2]{Color-color plots, showing the
  distribution of the SDSS comets. The resolved Jupiter Family Comets (JFC)
  are in black, while non-JFC resolved comets are in blue. 
  The boxes represent the range of 2 standard deviations for the
  resolved comet distribution (red),
  and Jupiter Trojans (green) from \cite{2007MNRAS.377.1393S}. $a_{pc}$ is the principle
  component color in the MOC defined as $a_{pc}  = 0.89(g-r)+0.45(r-i)-0.57$
  (see \citealt{2001AJ....122.2749I}).  The two unresolved JFCs are in
  orange.}
\label{fig_colorcolorbox} 
\end{figure}

Extending on the work of \cite{2010Icar..205..605S} we present the
SDSS $u-g$, $g-r$, $r-i$, and $i-z$ colors for the active comets
observations in Table \ref{table_color}.  Figures \ref{fig_color_hist}
and \ref{fig_colorcolorbox} illustrate the uniformity in the colors of the active
comets.  Active comets occupy a very narrow
distribution in color space, exhibiting the median colors of
$u-g:$ $1.57 \pm 0.21$, $g-r:$ $0.57 \pm 0.05$, $r-i:$ $0.22 \pm 0.07$,
and $i-z:$ $0.09 \pm 0.07$.   To aid in comparison to other Solar
System bodies observed in the SDSS, we also calculate the SDSS asteroidal
principal component color  ``$a_{pc}$,''  defined as $a_{pc} =
0.89(g-r)+0.45(r-i)-0.57$ \citep{2001AJ....122.2749I}; the
median $a_{pc}$ is $0.04 \pm 0.06$.  This value lies between the modes for the $a_{pc}$ color distribution of main-belt asteroids (dominated by S and C types), and within uncertainty equal to the median value measured for the Jovian Trojans \citep{2007MNRAS.377.1393S}.  

When these colors are transformed into BVRI photometric colors \citep{2007AJ....134..973I} they are shown be consistent with previous color measurements of active comets \citep[see][and references within]{2010Icar..205..605S} yeilding mean colors of B$-$V: 0.74, V$-$R: 0.44, and R$-$I: 0.58.  These colors of active comets are slightly bluer (less than $1-\sigma$) in B$-$V and V$-$R, and redder (less than $2-\sigma$) in R$-$I ($-0.09$ in B$-$V, $-0.06$ in V$-$R, and $+0.12$ in R$-$I) than the colors found for cometary nuclei by \cite{2009Icar..201..674L}.

\subsubsection{Surface Brightness Profiles}\label{sec_SBP}
For a simple steady-state coma, the surface
brightness should follow a $\rho^{-1}$ relation \citep{1987ApJ...317..992J}, with $\rho$ being
the projected linear distance from the nucleus.  A plot of surface brightness
(mag arcsec$^{-2}$) against the log of angular size, should therefore show
a logarithmic gradient $m=-1$.  \cite{1987ApJ...317..992J} observed this relation 
and showed that when the effects of radiation pressure 
are taken into account, the slope can become steeper, ($m= -1.5$).  
Generally most measurements of the surface brightness profiles
of comets tend to fall in the range of $-2 < m < -1$
\citep[][for example]{1987ApJ...317..992J, 1999A&A...349..649L}.  For
comets that show profiles steeper than $m=-1.5$ additional factors
such as grain fading, may play a role.   

\begin{figure}[p]
\begin{center}
\includegraphics[width=1.0\textwidth]{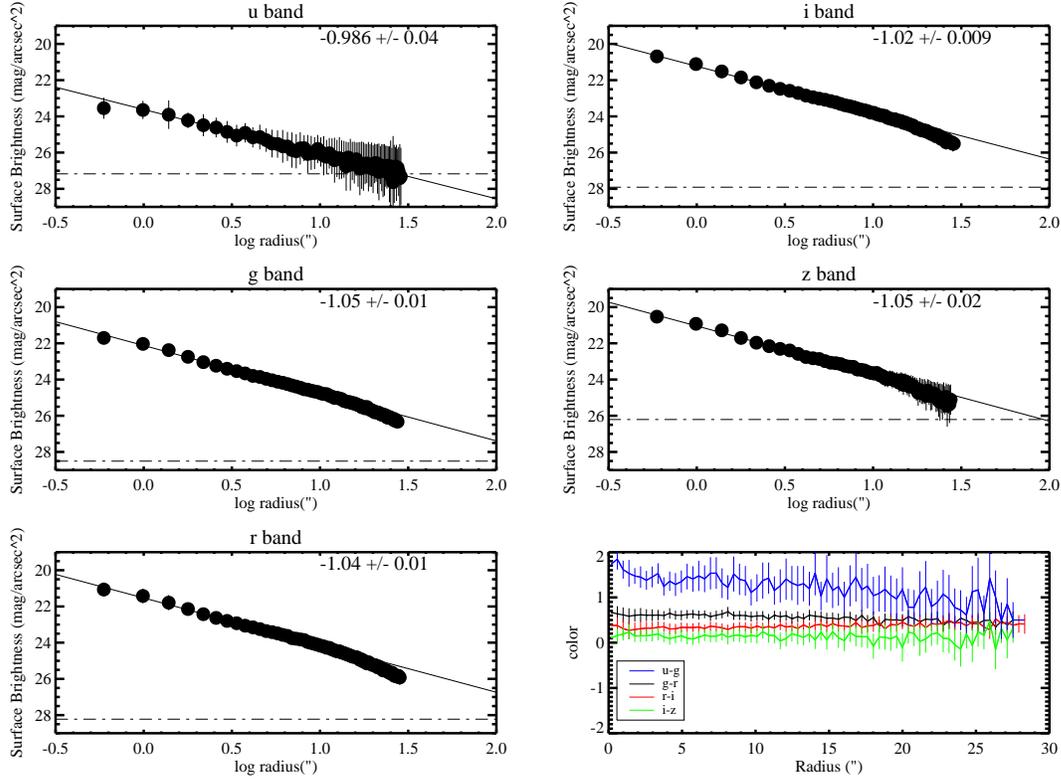} 
\end{center}
\caption[Surface brightness profile for comet
  67P/Churyumov-Gerasimenko in 5 bands] {Surface brightness profile of comet
  67P/Churyumov-Gerasimenko in $u,g,r,i,z$, along with the weighted
  least squares fit for the slope (solid black line, with value in
  each panel's upper right).  One count above sky would give the
  surface brightness shown by the horizontal dashed lines. The bottom
  right panel shows the local $u-g, 
  g-r, r-i$ and $i-z$ colors with respect to radius.  In all cases
  error bars are 1-$\sigma$. The color remains consistent
  with respect to radius until the local surface brightness becomes
  close to sky values (most prominently seen with $u-g$ in blue)}
\label{fig_CG} 
\end{figure}

The slopes of the observed active SDSS comets' surface brightness profiles were fit
using a weighted least-squares method independently in all five SDSS
bands; we report these values for all resolved comets in Table
\ref{table_slope}.  Figure \ref{fig_CG} shows the surface brightness
profiles and fits in all five bands for comet 67P/Churyumov-Gerasimenko,
all of which are fit by $m \sim -1$.  In addition, the
colors are constant over the extent of the comet's profile. 

\begin{figure} [p]
\begin{center}
\includegraphics[width=0.5\textwidth]{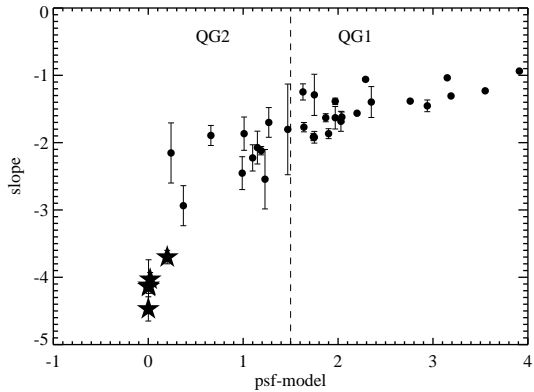} 
\end{center}
\caption[Plot of surface brightness profile slope vs. $r$ band
  PSF$-$model magnitude]{The average slope of the $u,g$,and $r$ band surface brightness
  profiles plotted against the PSF$-$model magnitudes of the active
  comets.  We divide the sample into two ``quality groups'' (QG1 and
  QG2) at PSF$-$model = 1.5.  For the more resolved QG1 comets the trend
  is fairly flat, and is characterized by generally small errors in
  the slope fits.  The less resolved QG2 comets trend toward steeper
  slopes, a trend that terminates with slopes fit to point-sources, as
  shown by the four stars in the plot.  The four objects near PSF$-$model = 0 
  (star symbols) are field stars
  whose surface brightness profiles were extracted and fit in the same
  manner as the comets.}
\label{fig_psf-model} 
\end{figure}

To express our confidence level in the measured surface brightness
profiles, we have divided our comets into two ``quality groups''  based
on the value of the difference between their PSF and model magnitudes.  
Comets that have small PSF$-$model magnitude values are less well resolved, 
and the accuracy of the fits
to these surface brightness profiles are not as robust as for better
resolved comets.  Figure \ref{fig_psf-model} shows the trend of the mean
slope (the mean of the $g,r$ and $i$ slopes) as a function of PSF$-$model 
magnitude, and demonstrates a break near
PSF-model = 1.5. At this point, the less resolved comets begin a trend toward
steeper slopes that extends all the way down to point-sources (PSF-model $\sim$ 0).  These
point sources (star symbols in Figure \ref{fig_psf-model}) are stars
from the comet fields of similar magnitude and were fit by
the same slope fitting routine used for the surface brightness
profiles of the comets. Since it is not clear which of the comets are less
resolved due to low activity versus those due to the resolution limit
of SDSS, we will use a value of PSF$-$model magnitude difference of
1.5 to define
two ``quality groups'' of surface brightness profile slopes.  The well
resolved comets (PSF$-$model $>$ 1.5) are placed in QG1 and the less resolved
comets (PSF$-$model $<$ 1.5) in QG2

\begin{figure} [p]
\begin{center}
\includegraphics[width=0.7\textwidth]{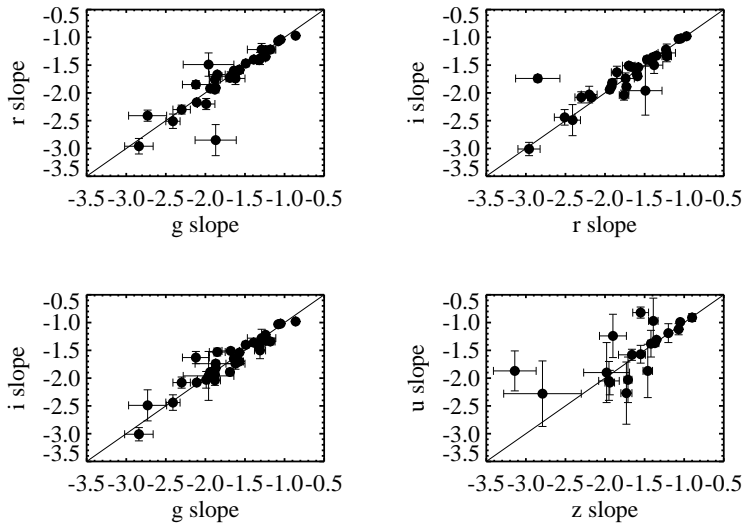} 
\end{center}
\caption[Correlation of surface brightness profile slope between filters]{Comparison 
between the best fit slopes of the surface brightness profiles for the SDSS filters.
The solid lines are the 1:1 relations.  Trends between
  the $g,r$ and $i$ band slopes are nearly 1:1, while there is
  more scatter in the $u$ and $z$ band trends.  The plotted
  points carry 1$\sigma$ error bars.}
\label{fig_slopeslope} 
\end{figure}

For several comets, the $u$ and $z$ bands do not have high enough surface
flux density to allow for robust fits to be made, and in those cases
no fit is given in Table \ref{table_slope}.  The $g,r,$
and $i$ bands are generally self consistent for a given comet.  
This trend of similar slopes from the $g,r,$ and $i$ band fits
is illustrated in Figure \ref{fig_slopeslope}.  The solid lines trace
a 1:1 relationship.  The fits to the $g,r,$ and $i$ band slopes trace each other well, while
the fits to the $u$ and $z$ bands show more scatter.  The
lower quality of fits in the $u$ and $z$ bands is due largely to the
surface flux density not being large enough compared to the sky to
allow for rigorous fits.  This results from a combination of the comets
themselves being fainter in the $u$ band, and the lower quantum
efficiency of the SDSS camera for the $u$ and $z$ filters
\citep{1996AJ....111.1748F, 2002AJ....123..485S}.

These five-band surface brightness profiles reveal that not only are the 
profile slopes consistent across a wide range of wavelengths, as seen
in Figure \ref{fig_slopeslope} and Table \ref{table_slope}, but
changes in the profile slope are also consistent from one band to the next.
\cite{1987ApJ...317..992J} predicted a transition to a steeper
profile slope at a distance from the nucleus where solar radiation
pressure becomes dominant.  Several bright comets
observed by the SDSS demonstrate this behavior, as illustrated by comet
67P/Churyumov-Gerasimenko in Figure \ref{fig_CG}.  The slope
transition is seen at the same distance (approximately $6\times10^5$km 
in the case of 67P) for a given comet across all
filters where the flux density is significantly above the background.

\subsubsection{Upper Limits to Cometary Radii}
Upper-limit estimates of the radius of the cometary nuclei may be
determined by using the PSF fitted 
magnitude for the resolved comets. We use the measured PSF magnitude from the 
SDSS photometric pipeline \citep{2002AJ....123..485S}, which involves 
sync-shifting the image so that it is exactly centered on a pixel, and then 
fitting a Gaussian model of the PSF to it, with additional corrections applied 
in order to take into account the full variation of the PSF across the field.
It would not be useful to take
flux from a large aperture for these determinations, as the light from
the coma is vastly more luminous than that from the nucleus. While
the detected flux in the PSF-magnitude does indeed
come from light interacting with the coma, the PSF-magnitude represents
the most light that could be theoretically coming from a point source in
the region of the nucleus.  In a sense this is the same assumption
that is made during the calculation of radii from photometry of
unresolved comets.  An unresolved comet does not show a coma at the
level of photometric resolution, and so the PSF photometry of the
object is taken to be the light reflecting from the nucleus, even
though that light may in fact be from an unresolved coma, rather than
the nucleus itself.  

Keeping in mind these assumptions, one can take the PSF$-$magnitude
measurement in the $r$ band as a nuclear upper limit, and calculate the absolute magnitude
$r_{(1,0)}$: the $r$ band
magnitude the comet would have with a geocentric and heliocentric
distance of 1 AU and viewed at a phase angle of zero degrees (this is a
physically impossible geometry). So for an observed $r$ band magnitude
\begin{equation}\label{e_rmag1}
r_{(1,0)} = r - 5 \log[R \Delta] - \phi(\alpha)
\end{equation}

where $R$ and $\Delta$ are the heliocentric and geocentric distances
in AU, $\alpha$ is the observed phase angle, and $\phi(\alpha)$ is the phase
function. This absolute magnitude is constructed in the same way using the 
SDSS $r$ band magnitude as is
the standard IAU asteroid $H$ magnitude with the Johnson $V$ band
magnitudes (for reference $r_{(1,0)} \sim H - 0.2$).
This absolute magnitude can be written in terms of the comet's
diameter, $D$ and albedo, $A$ \citep{2001AJ....122.2749I}.
\begin{equation}\label{e_rmag2}
r_{(1,0)} = 17.9 - 2.5\log(\frac{A}{0.1}) - 5 \log (\frac{D}{1km})
\end{equation}
By assuming a standard albedo of $0.04$ and adopting a linear phase
function, $\phi(\alpha) = \beta \alpha$, with $\beta=0.035$ mag deg$^{-1}$,
these two equations can be used to 
transform the PSF $r$-band magnitudes into upper limits for the sizes
of the comet nuclei. The upper radii limits derived in this manner
(Table \ref{table_derived}) are comparable to JFC radii in the literature, in particular the 
list of Jupiter Family Comet size estimates made by \cite{2006Icar..182..527T}, 
who also assume an albedo of 0.04.  The trend is
that the SDSS upper limits tend to be larger than those published by
\cite{2006Icar..182..527T} by about 75\%, particularly for the
comets for which \cite{2006Icar..182..527T} express high levels of confidence in
their estimates.

\subsubsection{Dust Production Rates}

The dust production rate of comets gives insight into the current status, and
if measured over time, the evolution of the comet's activity. Dust
production rates for the observed comets may be characterized from their
photometry through the quantity A$f\rho$ defined by
\cite{1984AJ.....89..579A}.  Where A is the dust grain albedo, $f$ the filling factor 
within the chosen aperture,  and $\rho$ the linear radius corresponding to the 
aperture. A$f\rho$ can be computed directly from observable quantities
\begin{equation}
Af\rho (\textrm{cm})= \frac{(2 \Delta R)^{2}}{\rho} \frac{F_{comet}}{F_\odot}
\end{equation}

Here $\Delta$(cm) and $R$(AU) are the geocentric and heliocentric
distances, $\rho$(cm) is the linear radius of the photometric aperture, and
$F_{comet} / F_{\odot}$ is the ratio of the observed comet flux
to that of the Sun. Empirical comparisons indicate a linear
relationship between the  A$f\rho$ value and the dust production rate,
with A$f\rho$ (measured in units of 1000 cm) being roughly equal to the modeled dust
production rate in metric tons per second \citep{1995Icar..118..223A}. 

For an idealized steady-state coma, the surface
brightness profile follows a $\rho^{-1}$ profile.  Under that coma model,
A$f\rho$ is independent of aperture, but since real comae deviate from
$m=-1$ it is necessary to define the radius at which A$f\rho$ is to be
measured. Rather than assuming an arbitrary radius we take a cue from galaxy photometry
\citep{1976ApJ...209L...1P} and define the radius, $\rho$ 
at which A$f\rho$ is calculated  as the radius at which the local
surface brightness value is 10\% of the mean enclosed local
surface brightness value, $\frac{SB(\rho)}{mean[SB(\le\rho)]} =
10\%$.  This criterion results in A$f\rho$ being measured at
approximately $10^{3} - 10^{5}$km from the center of the comet, and is consistent
with radii chosen in the literature for comets over a wide range of observational 
distances (cf. \citealt{LIII} and \citealt{2007MNRAS.381..713M}).

While under idealized conditions A$f\rho$ is independent of aperture
(for a $\rho^{-1}$ profile), and wavelength (for a grey dust approximation), it is
dependent on the observational phase angle.  By adopting a simple
phase function for single dust particles from
\cite{1981ESASP.174...47D},  the measured A$f\rho$ values may be
corrected to A$(0)f\rho$, the expected value if the comet had been at
zero phase angle during the observation.
As most reported A$f\rho$ data in
literature are not phase corrected, we will generally refer to our
uncorrected A$f\rho$ values, unless specifically referencing the
phase-adjusted A$(0)f\rho$. Table \ref{table_derived}
lists the Af$\rho$ and A$(0)f\rho$ measurements for the active comets.

\subsection {Unresolved Comets}\label{s_unresolved}

Four comets in our sample are unresolved by SDSS.  Two of these, 
19P/Borrelly and 113P/Spitaler, are dynamically classified as Jupiter
Family Comets.  174P/Echeclus is the Centaur 60558 Echeclus that has
been observed to have activity \citep{2006IAUC.8656....2C}, and
176P/LINEAR is a main belt comet, previously designated as main 
belt asteroid 118401 LINEAR  \citep{2006Sci...312..561H}. All
four of these unresolved comets are in 
the SDSS MOC; 124P and 126P are correctly identified and matched to
their asteroid designation, while 19P and 113P, which are in comet 
but not asteroid databases, are unmatched sources in the
SDSS MOC.  
There are likely to be more unresolved
comets to be found in the SDSS MOC, but identifying and retrieving 
these objects is difficult due to non-gravitational forces causing them to deviate from projected
positions on the sky.

The  $u-g$, $g-r$, $r-i$, and $i-z$ colors are presented in Table \ref{table_color}.  
Figure \ref{fig_colorcolorbox} shows that the unresolved JFCs,
19P/Borrelly and 113P/Spitaler, fall into the color space spanned by
the active comets. 

If we assume that these observations of unresolved comets are images of
the inactive cometary nucleus, we can make an estimate of the comet's
size assuming an albedo of 0.04. These are included in
Table \ref{table_derived}, and like the upper limits estimated for the
resolved comets, are in satisfactory agreement with previously
determined values. 

\section {Discussion}\label{disc} 

\subsection{Photometric Colors}\label{sec_dis_color}

\begin{figure}[p]
\begin{center}
\includegraphics[width=1.0\textwidth]{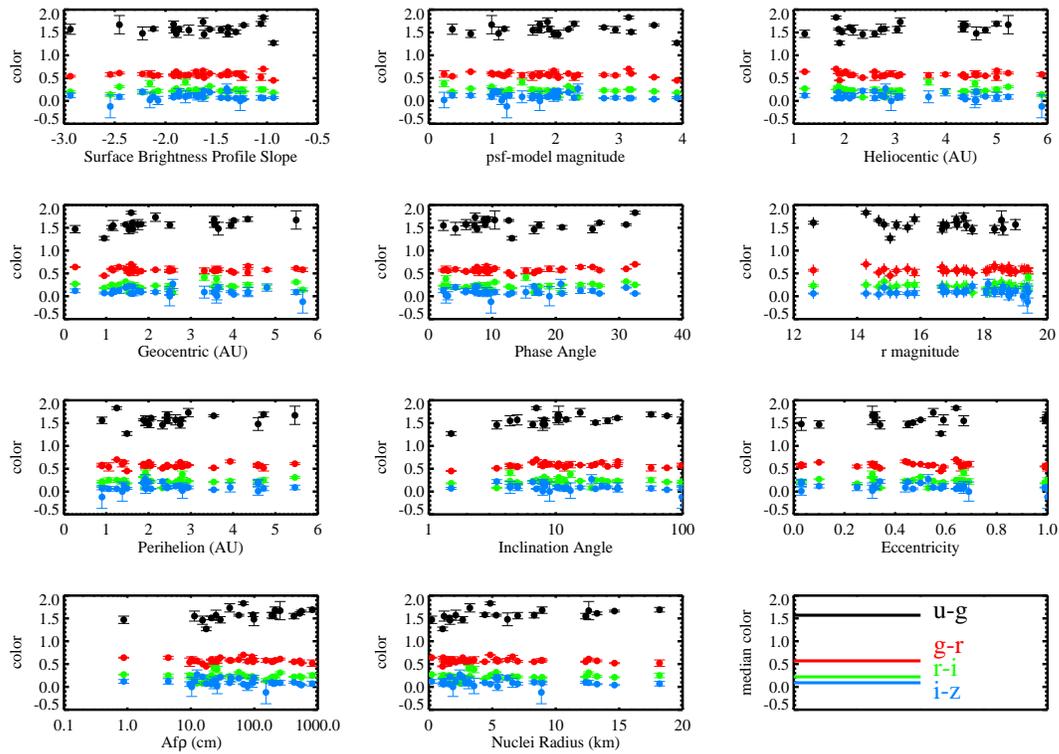} 
\end{center}
\caption[Comet colors compared to various parameters]{The SDSS colors
  of comets are compared against various
  parameters, the lower right plot has a key to the colors in the
  plot, as well as showing the median value for each color.  There is
  no statistically significant trend in color
  against any parameter.  The slopes of these trends have been fit,
  and are presented in Fig \ref{fig_constcol}.  }
\label{fig_rainbow} 
\end{figure}

\begin{figure}[p]
\begin{center}
\includegraphics[width=.75\textwidth]{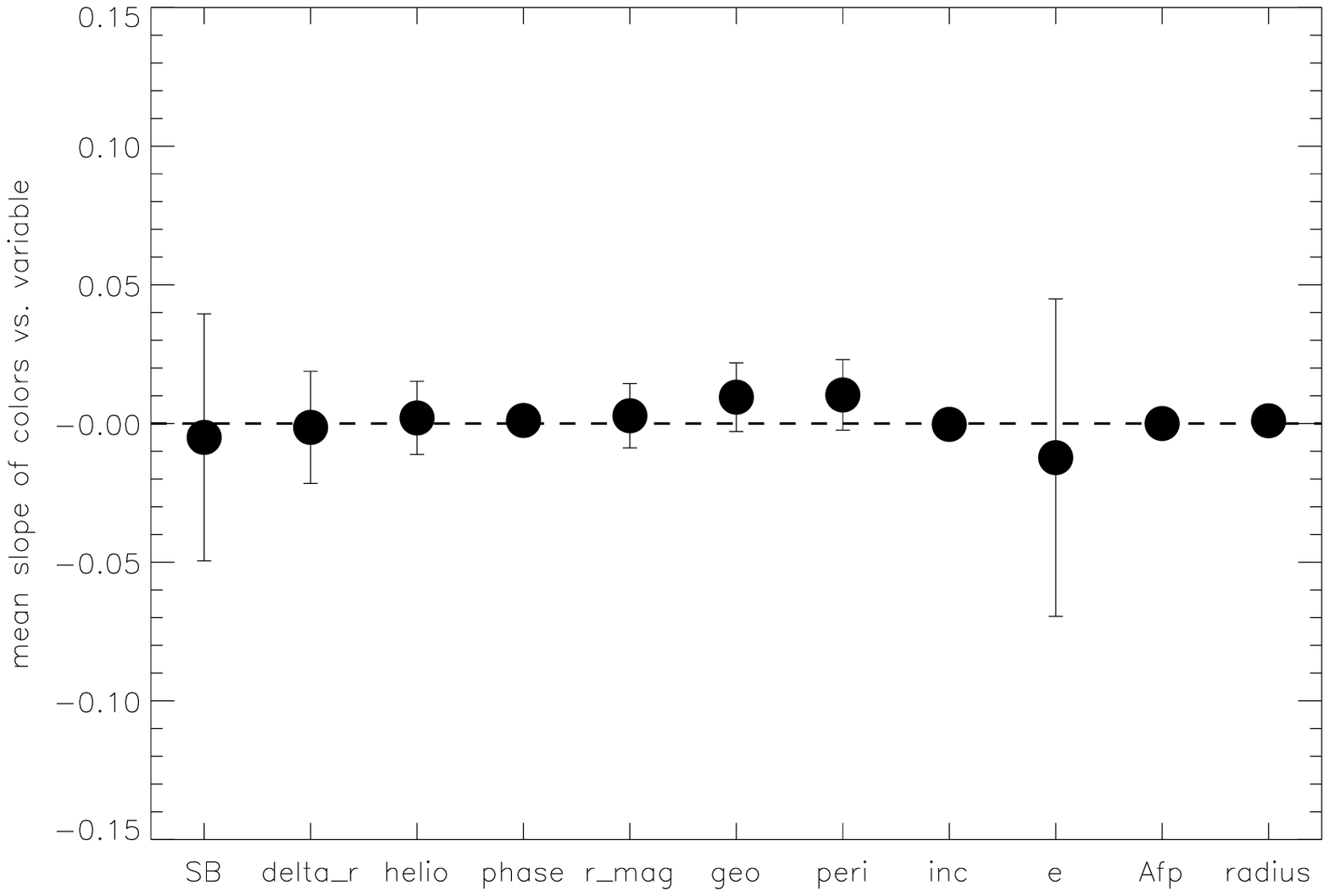} 
\end{center}
\caption[Comet colors are constant]{The plot of mean color gradient of
  SDSS colors of
  comets against various parameters shows no color dependence on
  them. These same parameters can be seen
  in Figure \ref{fig_rainbow}.  The abbreviations are:\\
-SB: Slope of the surface brightness profile\\
-delta\_r: The $r$-band PSF-model magnitude\\
-helio: The heliocentric distance in AU at the time of observation\\
-phase: The phase angle at the time of observation\\
-r\_mag: The $r$-band magnitude\\
-geo: The geocentric distance in AU at the time of observation\\
-peri: The perihelion distance in AU\\
-inc: The orbital inclination\\
-e: The orbital eccentricity\\
-A$f\rho$: The calculated A$f\rho$ value in cm\\
-radius: The calculated radius limit in km
}
\label{fig_constcol} 
\end{figure}

In light of the significant differences of observational geometry,
    physical parameters, and orbital type, all active comets have the
    same SDSS
    colors, which span a range of $\sim 6000$ \AA\ across the five
    filters.  The photometric ($u-g$, $g-r$,
$r-i$, and $i-z$) colors of the
active comets (Table \ref{table_color}) reveal that they are remarkably similar to each other,
particularly in $g-r$, $r-i$, and $i-z$, where the colors have
a scatter of $\le0.07$ magnitude.  Furthermore, this uniformity of color does not show
systematic variation with observational geometry (phase angle, geocentric
or heliocentric distance), measured quantities ($r$ band magnitude,
PSF$-$model magnitude, slope of the surface brightness profile), derived
quantities (nuclear radius, A$f\rho$), or orbital properties (orbital
inclination, eccentricity and perihelion distance). Figure
\ref{fig_rainbow} illustrates the dependence of all
four colors on each of these parameters.  The colors are consistent with no slope
 within $1\sigma$ against all parameters as
shown in Figure \ref{fig_constcol}.  The result of the colors of
active comets having no correlation with heliocentric distance is
consistent with the results of \cite{1988AJ.....96.1723J}. Considering that the dust 
in the coma is thought to be
    primordial solar system grains freed by the volatilization of ices,
    this result may indicate that
    such a color uniformity could be a common property of the
    primordial dust grains of the outer solar system.

The colors of the comets discussed here are strikingly
similar to that of the Trojans of Jupiter \citep{2007MNRAS.377.1393S} as 
shown in Figure \ref{fig_colorcolorbox}. Trojans also show similar 
colors to active comets in 5.2-38$\mu$m
observations \cite{2006Icar..182..496E}, which may indicate a link
between the surface chemistry and composition of comets and Trojans.

\begin{figure} [p]
\begin{center}
\includegraphics[width=.6\textwidth]{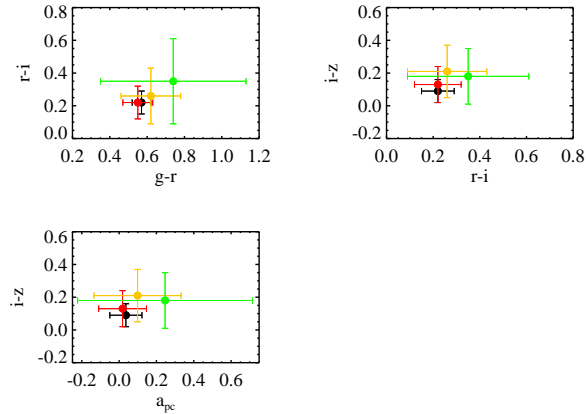} 
\end{center}
\caption[Color-color diagram of small body populations in SDSS]
	{Color-color diagrams of the various small body populations
	discussed in \ref{sec_dis_color}. The active comets are in
	black, the Trojans are in red, the Centaurs are in orange, and
	the TNO's are in green.  The error bars represent one standard
	deviation.}
\label{fig_colorcolor_pop} 
\end{figure}

To further characterize this color similarity we examine the gradient of the normalized
reflectivity, $S^{'}$, expressed as \% per 1000\AA.  The colors and
distribution of $S^{'}$ for the active comets are 
compared to those of a sample of 363 Trojans, 12
Centaurs, and 23 Trans Neptunian Objects (TNOs) from the SDSS.  The 
Trojan and Centaur samples were selected by
their orbital elements (and
positive cross-identification) from the SDSS MOC, and the TNOs were
found in SDSS Stripe 82 by \cite{becker}.  Figure
\ref{fig_colorcolor_pop} shows the range in color-color space
occupied by these populations, with the Trojans and active comets
being in close agreement.
The SDSS colors found in this work agree with those of
\cite{2007MNRAS.377.1393S}, and the $S^{'}$ for the active comets
resembles the $S^{'}$ distribution of JFC nuclei found by
\cite{2002AJ....123.1039J}. Note that Centaurs are extremely faint
objects with correspondingly higher photometric errors.

\begin{figure} [p]
\begin{center}
\includegraphics[width=.75\textwidth]{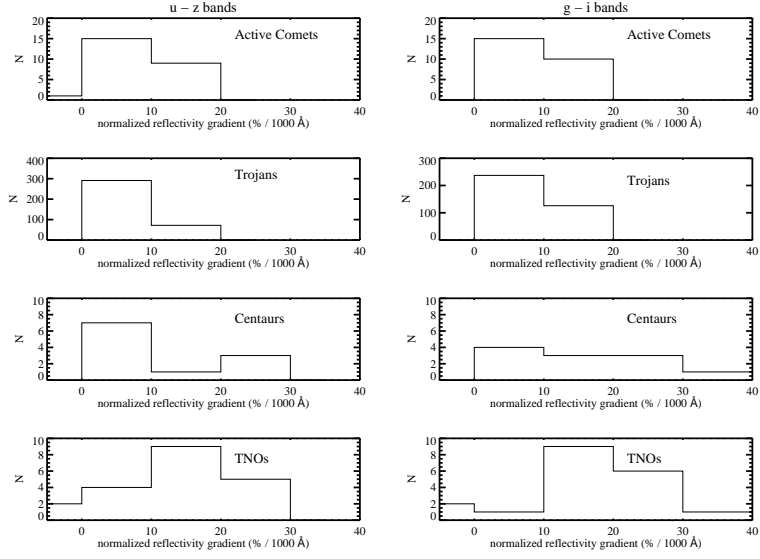} 
\end{center}
\caption[Histogram of $S^{'}$]
	{Histograms of $S^{'}$ for the various small body populations
	discussed in \ref{sec_dis_color}. }
\label{fig_sprime} 
\end{figure}

Histograms of these three populations are seen in Figure
\ref{fig_sprime}.  The faintness of the Centaurs and TNOs
makes the full five-band gradient less reliable.  The colors and
reflectivity gradients for these populations are
summarized in Table \ref{table_sprime}. By visual inspection,
these histograms confirm in $S^{'}$ what was seen in their colors:
the active comets and Trojans are remarkably similar.  To characterize
this similarity a
Kolmogorov-Smirnov test (KS-test) was performed on the $S^{'}$
distributions of the populations in question. Between the Trojans and
the Comets, the KS-test probability value was
0.16 with respect to the $S^{'}$ distribution across all five bands,
and rises to 0.81 for the $gri$ band, suggesting that the comets and
 the Trojans could have been drawn from the same population.  To
 reiterate, even though Trojans and Comets are two different types of object,
 with the light we see reflecting off of different sources (solid body
 vs. dusty coma), their colors are statistically indistinguishable.

\subsection{Nuclear Radii}\label{s_radii}

Estimates of absolute magnitude and nuclear size were made using equations \ref{e_rmag1} 
and \ref{e_rmag2}.  Despite the
measurements of nuclear radii for the active comets being the result of
accepting an upper limit from the SDSS PSF magnitude, the radii
presented here agree with the measurements in literature as
shown in Table \ref{table_derived}.

Even for targeted comet surveys, it is difficult to be certain that all of the 
flux seen is really light that has
interacted with the surface of the comet;  in some cases there may be
low-level activity below the SDSS resolution limit.  Certainly there are 
methods of obtaining highly accurate radius measurements, the 
best obviously being \emph{in situ} spacecraft
encounters with comets.  Only five comets, 1P/Halley
\citep{1986Natur.321..313R}, 19P/Borrelly \citep{2002Sci...296.1087S},
81P/Wild-2 \citep{2004Sci...304.1764B}, 9P/Tempel-1
\citep{2005Sci...310..258A}, and 103P/Hartley-2 \citep{2011AAS...21730608A} have to date had observations of the nucleus by remote
spacecraft.

 \cite{1995A&A...293L..43L} outline how to use
the high spatial resolution afforded by the Hubble Space Telescope
to disentangle the nucleus signal from the inner coma.  Highly accurate
results may also be produced if one has access to simultaneous observations of
the comet at both visible and infrared wavelengths
\citep{2002Icar..156..442L}.  Most
cometary observations, however, must make use of similar methods as
employed here.  For resolved comets the coma must be algorithmically
removed, often either through the modeling of the coma and
subtraction, or through the use of small aperture photometry or PSF
model fitting.  Even unresolved comets
must be assumed to be inactive for observation of the radius. 
Complicating the matter further is the
non-uniformity of cometary albedos: while comets are dark overall, the
albedo can vary comet-to-comet, and even across the surface of
an individual one \citep{Buratti200416}.
Being small collisionally-shaped bodies, they can have 
axial, and even tri-axial shapes.  This feature presents an even larger problem as very
few comets have had their nuclear light curve accurately measured,
(only a few dozen comets have robustly measured nucleus observations) and
many of those show dramatic asymmetries (e.g. comet
19P/Borrelly with a primary axis ratio of 2.5:1 \citep{Buratti200416}).

\begin{figure} [p]
\begin{center}
\includegraphics[width=.75\textwidth]{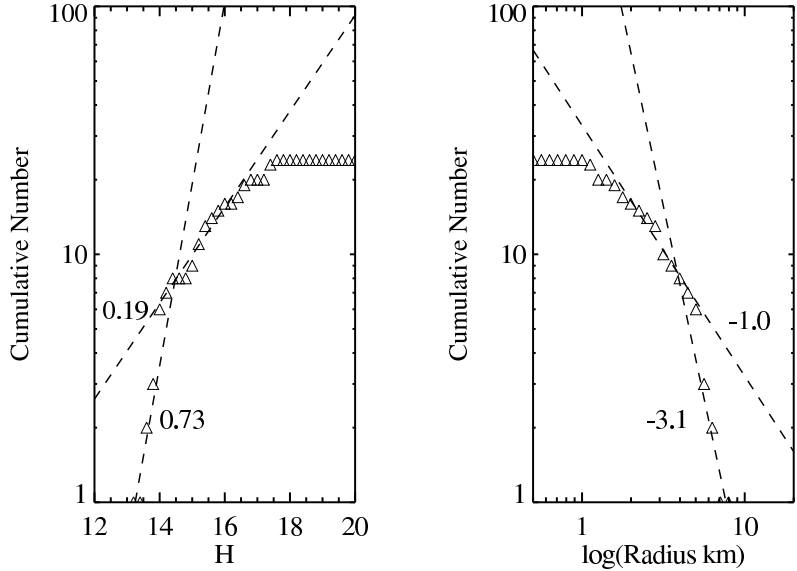} 
\end{center}
\caption[Cumulative luminosity function and size distribution for SDSS comets]{\textbf{Left:} the Cumulative Luminosity Function for the
  Jupiter Family Comets 
  in this paper, with power law fits.  The whole population may be fit with a gradiant of $0.49 \pm 0.05$. Fitting with a broken power law at $H\sim 14.5$ gives gradients of $0.73 \pm
  0.08$ and $0.19 \pm 0.03$ for the two fits, suggesting the faint population may be shallower than indicated by the single fit.  \textbf{Right:} the Size
  Distribution Function for the same sample, assuming an albedo of
  0.04. There is a break between the 
  two measured slopes at $\sim 4$km.  The gradients are $-3.1 \pm
  0.3$ and $-1.0 \pm 0.1$ for the broken fit. Fitting the whole range
  gives a fit of $-2.4 \pm 0.2$.}
\label{fig_CSD} 
\end{figure}

Although the SDSS resolution is not sufficient for direct measures of the 
radii they do have the advantage that all observations are made and 
reduced by a well calibrated and highly characterized process.  
Thus, the large number of comets observed by the SDSS allows a statistical study of
the JFC population, with confidence that
the data set is self consistent.  To better compare these results with
previous cometary studies, we have used the estimated radii upper limits of the
comets to calculate (assuming an albedo of 0.04) the more
commonly used absolute $H$ magnitude.
Figure \ref{fig_CSD} shows the Cumulative Luminosity Function (CLF - the
number of JFCs more luminous than a given $H$), and the
Cumulative Size Distribution (CSD) for the JFCs.  These distributions 
may be fit with power laws, with the CLF best fit
by an exponent of $0.49 \pm 0.05$, and the CSD by $-2.4 \pm
0.2$ for $ H < 18$. These give the functional form of the distributions as

\begin{equation}
\textrm{CLF: } N(<H) \propto 10^{0.49 H}
\end{equation}
\begin{equation}
\textrm{CSD: } N(<r) \propto r^{-2.4}
\end{equation}

for a given absolute magnitude $H$ and radius $r$.  These values are
in line with the results of previous studies, which are summarized in Table \ref{table_CLF}. 

We find that these distributions may also be fit with a
broken power law, with breaks at $H=14.5$ in the CLF, and at radius = 4km
in the CSD (Figure \ref{fig_CSD}).  For the CLF the gradients 
are $0.73\pm0.08$ for $H<14.5$ and $0.19\pm0.03$ for $14.5<H<18$ 
yielding $-3.1 \pm 0.4$ and $-1.0 \pm 0.1$ for the CSD. 
This is similar to the CLF measured by \cite{LIII} when
considering their entire sample (both resolved and unresolved) of 
comets ($0.53 \pm 0.04$ and $0.22 \pm 0.02$).  We find that our shallow
power law for smaller comets are is agreement, but find that our sample supports a steeper fit for the larger comets.  Again these slopes are derived from both radii found for unresolved comets, and upper limits placed on the nuclear radii of resolved comets.  While the parameters of this broken power law fit are not strongly constrained they do suggest that that faint comet population could be much shallower than indicated by a single power law fit.

Table \ref{table_popCLF} compares our results for the Jupiter Family
Comets with those of other small body populations.  As is the case for many of these
populations, the observed CLF power law exponent is similar to
0.5, predicted by \cite{1969JGR....74.2531D} for a model based on an
equilibrium cascade of self-similar collisions.  The large number of
Main Belt Asteroids observed by the SDSS allows their CLF to be analyzed
family-by-family, and analytically fit to a function that accounts for
a change in the power law slope \citep{2008Icar..198..138P}.  With the
multitude of observations on other small body populations expected
from the next generation of sky surveys \citep[e.g. Pan-STARRS and
  LSST,][]{2002SPIE.4836..154K, 2008arXiv0805.2366I}, it is expected
that similar detailed analysis could be performed on the less
well-sampled populations including Jupiter Family Comets.

\subsection{A$f\rho$}
The A$f\rho$ values obtained for the active comets observed by the
SDSS cover a range of three orders of magnitude 0.88 cm
$\leq$ A$f\rho$ $\leq$ 817.68 cm.  Table \ref{table_Afp} compares the Af$\rho$ values
determined in this paper with those from other studies. They are consistent 
with other measurements;  in particular
67P/Churyumov-Gerasimenko (the ESA Rosetta Mission target) has a well
sampled A$f\rho$ over the comet's orbit.  Our A$f\rho$ value for this comet is
consistent with that found at the same phase angle by \cite{2010arXiv1001.3010A}. 

\begin{figure} [p]
\begin{center}
\includegraphics[width=.75\textwidth]{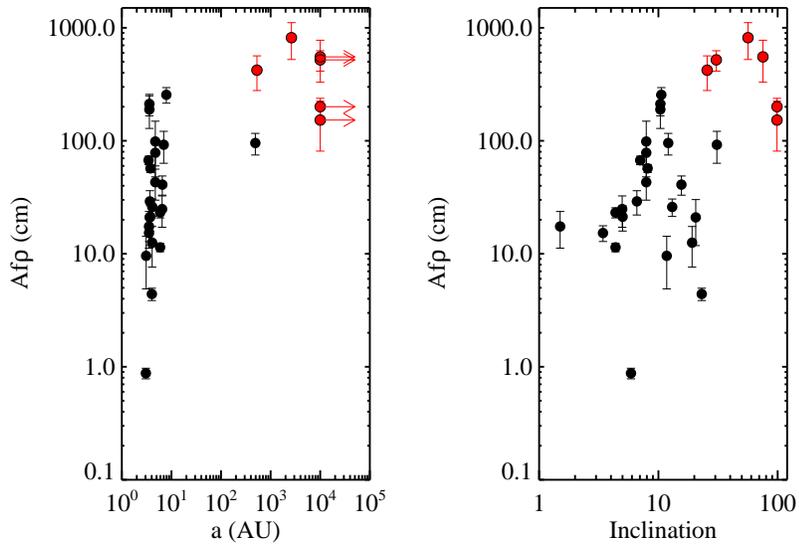} 
\end{center}
\caption[LPC have higher A$f\rho$ than JFC]{Calculated A$f\rho$ values are
  compared on the left to the semi-major axis (a) and on the right to
  orbital inclination of the comets.  Jupiter Family Comets are in
  black.  The non-JFC population shows statistically larger A$f\rho$ values.  Note: LPCs plotted at $10^4$AU are
  comets on unbound orbits, indicated by arrows.
 }
\label{fig_afpLPC} 
\label{lastfig}
\end{figure}

Using A$f\rho$ as proxy for dust production, these values show a trend of the
non-JFC comets having significantly higher dust production than do JFC 
comets.  This supports the suggestion that Long Period Comets being
recently introduced to the inner solar system, are not
as surface-volatile depleted as a Jupiter Family Comet that would have
presumably made multiple perihelion passages (cf Figure \ref{fig_afpLPC}).  Other factors such as a difference in the size distribution or albedo of the dust between these two populations could also produce a change in A$f\rho$.

\section {Summary}\label{sum}

We have analyzed data for 31 comets observed in the SDSS, which provides
accurately calibrated and measured multi-wavelength photometry in five bands.
These comets span a
wide range of heliocentric distances, observational, and orbital parameters and all have been observed with the same instrument and processed with the same software. This photometry has been used to make
measurements of the colors, sizes, surface brightness profiles and rates
of dust production (in terms of the A$f\rho$ formalism).  Our main results are as follows. 

\begin{enumerate}

\item  Despite the variety of cometary parameters, the
distributions of  photometric colors are extremely narrow ($0.57 \pm 0.05$ in $g-r$ for
    example), and statistically indistinguishable from those of
    Jupiter Trojans.  The comets exhibit \emph{no}
    correlation between color and physical, dynamical, or
    observational parameters, as seen in Figure \ref{fig_constcol}.  Additionally 
    the surface brightness profile for each comet is found to be invariant 
    with wavelength in the optical. 

\item The uniform, red color of the comets indicates that the light from
the dust in the coma is complex, no showing any systematic variation with observational geometry, dynamics or physical properties of the comets, and can not be
explained through simple scattering (e.g. a simple $1/\lambda^n$ optical depth dependance with $n\sim4$). 
The visible light observed here is light
scattered by the larger particles in the coma, either conglomerates of
small grains, or larger, macroscopic rocks.  The similarity in reflectance
to solid surfaces (both cometary nuclei, and Jupiter Trojans) may
indicate that the material responsible for the dark, reddish
appearance of these solid objects is also present in the coma and
dictates its optical properties.

\item We find that the cumulative
    luminosity function for the Jupiter Family Comets can be
    fit by a power law of the form $N(<H) \propto 10^{(0.49
    \pm 0.05) H}$, with evidence of a broken power law with a exponent of
    ${0.73\pm0.08}$, transitioning to ${0.19\pm0.03}$ at $H\sim14.5$, suggesting a shallower gradient for the faint population.
    These results are consistent with distributions of small bodies,
    both of JFCs, and other Solar System populations. 
    
\end{enumerate}

The analysis presented here is relevant to upcoming large-sky surveys
such as the Dark Energy Survey \citep{22007.BASS.2009}, Pan-STARRS
\citep{2002SPIE.4836..154K} and the Large Synoptic Survey 
Telescope \citep[LSST]{2008arXiv0805.2366I}.  The results of this work
represent a ``snap-shot'' survey of comets, observing a sample of
comets at a single point in their orbit.
\cite{2010Icar..205..605S} estimate that $10^3$ to $10^4$ comets will be observed in
a survey such as LSST, with a limiting magnitude of $r\sim 24.5$.
Further, over the course of 10 years LSST will observe each comet
repeatedly.  Based on the results of a simulated LSST observational
cadence \citep{2008arXiv0805.2366I}, a JFC such as 70P/Kojima would be
observed 300-400 times, spanning its whole heliocentric range.  Such
observations mean that not only may the analysis presented here be
done for a vast number of comets, down to even fainter limiting
magnitudes, but the time evolution of each of these parameters may be
studied over the comet's orbit.

\section*{Acknowledgments}

The authors would like to thank Dr. Yanga Fern\'{a}ndez for
providing an up to date list of orbital elements for periodic comets, and our two anonymous reviewers for many insightful comments that helped improve the final version.

M.S. would like to acknowledge support from the Brinson Foundation grant in aid of Astrophysics Research to the Adler Planetarium \& Astronomy Museum.

Z.I. acknowledges support by NSF grant AST-0551161 to LSST for design and development activity, and by the Croatian National Science Foundation grant O-1548- 2009.Ê

M.J. wishes to acknowledge the support for this work provided by NASA
through Hubble Fellowship grant \#HF-51255.01-A awarded by the Space
Telescope Science Institute, which is operated by the Association of
Universities for Research in Astronomy, Inc., for NASA, under contract
NAS 5-26555.

D.P.S. was supported in part by National Science Foundation grant AST 06-07634

IDP3 was developed by the Programming Group for the NICMOS IDT Project
(NASA Grant NAG 5-3042, Rodger I. Thompson, P.I.) as a tool for image
manipulation and visualization. Many people have contributed to the
development of IDP3.\\  The IDP3 website can be found at
http://mips.as.arizona.edu/MIPS/IDP3/.

    Funding for the SDSS and SDSS-II has been provided by the Alfred
    P. Sloan Foundation, the Participating Institutions, the National
    Science Foundation, the U.S. Department of Energy, the National
    Aeronautics and Space Administration, the Japanese Monbukagakusho,
    the Max Planck Society, and the Higher Education Funding Council
    for England. The SDSS Web Site is http://www.sdss.org/. 

    The SDSS is managed by the Astrophysical Research Consortium for
    the Participating Institutions. The Participating Institutions are
    the American Museum of Natural History, Astrophysical Institute
    Potsdam, University of Basel, University of Cambridge, Case
    Western Reserve University, University of Chicago, Drexel
    University, Fermilab, the Institute for Advanced Study, the Japan
    Participation Group, Johns Hopkins University, the Joint Institute
    for Nuclear Astrophysics, the Kavli Institute for Particle
    Astrophysics and Cosmology, the Korean Scientist Group, the
    Chinese Academy of Sciences (LAMOST), Los Alamos National
    Laboratory, the Max-Planck-Institute for Astronomy (MPIA), the
    Max-Planck-Institute for Astrophysics (MPA), New Mexico State
    University, Ohio State University, University of Pittsburgh,
    University of Portsmouth, Princeton University, the United States
    Naval Observatory, and the University of Washington.

\label{lastpage}

\newcommand{\aj}{AJ }
\newcommand{\apjs}{ApJS }
\newcommand{\iaucirc}{IAU Circ. }
\newcommand{\aap}{A\&A }
\newcommand{\apj}{ApJ }
\newcommand{\mnras}{MNRAS }
\newcommand{\jgr}{JGR }
\newcommand{\nat}{Nature }

\bibliographystyle{apj}
\bibliography{comet2}


\clearpage



{
\renewcommand{\baselinestretch}{1}
\small\normalsize

\begin{table} [p]
\begin{center}
\tiny
\begin{tabular}{|l|c|c|c|c|c|}
\hline \hline
Comet	&	$r$			&	PSF$-$model	&	$\Delta$	&	R	&	Phase	\\  \hline
$\dagger$19P/Borrelly	&	21.35	$\pm$	0.05	&	-	&	3.685	&	4.623	&	5.1	\\  
30P Reinmuth	&	14.85	$\pm$	0.01	&	2.20	&	1.459	&	1.884	&	31.1	\\  
46P/Wirtanen	&	18.84	$\pm$	0.02	&	1.63	&	1.641	&	2.598	&	7.9	\\  
47P/Ashbrook-Jackson	&	20.57	$\pm$	0.05	&	0.24	&	3.606	&	4.572	&	2.7	\\  
50P/Arend	&	18.13	$\pm$	0.03	&	2.35	&	2.574	&	2.809	&	20.8	\\  
62P/Tsuchinshan	&	15.04	$\pm$	0.01	&	3.91	&	0.953	&	1.904	&	13.1	\\  
64P/Swift-Gehrels	&	19.23	$\pm$	0.04	&	1.75	&	2.497	&	2.917	&	19	\\  
65P/Gunn (1)	&	17.13	$\pm$	0.01	&	1.75	&	3.551	&	4.340	&	8.7	\\  
65P/Gunn (2)	&	17.15	$\pm$	0.01	&	1.90	&	3.559	&	4.338	&	8.9	\\  
67P/Churyumov-Gerasimenko	&	14.28	$\pm$	0.01	&	3.15	&	1.590	&	1.836	&	32.5	\\  
69P/Taylor	&	15.59	$\pm$	0.01	&	3.19	&	1.127	&	1.950	&	21.0	\\  
70P/Kojima	&	16.68	$\pm$	0.01	&	1.97	&	1.629	&	2.594	&	7.6	\\  
$\dagger$113P/Spitaler	&	21.05	$\pm$	0.05	&	-	&	2.332	&	3.320	&	2.3	\\  
129P/Shoemaker-Levy	&	18.07	$\pm$	0.02	&	2.03	&	2.495	&	3.051	&	17.0	\\  
146P/Shoemaker-Linear	&	18.78	$\pm$	0.02	&	1.27	&	1.301	&	2.023	&	23.9	\\  
158P/Kowal-LINEAR (1)	&	18.78	$\pm$	0.02	&	1.15	&	3.619	&	4.596	&	3.0	\\  
158P/Kowal-LINEAR (2)	&	18.60	$\pm$	0.02	&	1.10	&	3.650	&	4.597	&	4.2	\\  
158P/Kowal-LINEAR (3)	&	19.36	$\pm$	0.02	&	1.01	&	3.988	&	4.798	&	7.3	\\  
$\dagger$174P (60558 Echeclus)	&	21.20	$\pm$	0.08	&	-	&	14.350	&	15.167	&	2.27	\\  
$\dagger$176P  (118401 LINEAR)	&	20.27	$\pm$	0.03	&	-	&	2.374	&	3.256	&	9.2	\\  
P/2002 EJ57 (LINEAR)	&	18.99	$\pm$	0.01	&	0.37	&	1.752	&	2.720	&	5.9	\\  
2003 WY25	&	18.33	$\pm$	0.01	&	0.66	&	0.265	&	1.217	&	25.8	\\  
C/1999 F2 (Dalcanton)	&	15.81	$\pm$	0.01	&	2.29	&	4.344	&	4.996	&	9.3	\\  
C/2000 K2 (LINEAR)	&	16.82	$\pm$	0.01	&	1.74	&	3.950	&	4.776	&	7.2	\\  
C/2000 QJ46 (LINEAR) (1)	&	17.44	$\pm$	0.01	&	1.64	&	1.162	&	2.168	&	2.3	\\  
C/2000 QJ46 (LINEAR) (2)	&	19.39	$\pm$	0.04	&	1.47	&	3.318	&	3.657	&	15.3	\\  
C/2000 SV74 (LINEAR)	&	14.68	$\pm$	0.00	&	3.55	&	4.015	&	4.380	&	12.6	\\  
C/2000 Y2 (Skiff)	&	16.71	$\pm$	0.01	&	1.19	&	1.824	&	2.785	&	5.8	\\  
C/2001 RX14 (LINEAR)	&	12.62	$\pm$	0.01	&	2.76	&	1.619	&	2.103	&	26.9	\\  
C/2002 O7 (LINEAR) (1)	&	19.37	$\pm$	0.03	&	1.23	&	5.645	&	5.886	&	9.8	\\  
C/2002 O7 (LINEAR) (2)	&	15.24	$\pm$	0.00	&	2.94	&	2.504	&	3.041	&	17.4	\\  
P/1999 V1 (Catalina)	&	17.36	$\pm$	0.01	&	1.87	&	2.164	&	3.099	&	7.3	\\  
P/2002 T5 (LINEAR)	&	18.33	$\pm$	0.02	&	1.97	&	4.351	&	5.002	&	9.1	\\  
P/2004 A1 (LONEOS)	&	18.55	$\pm$	0.01	&	0.99	&	5.226	&	5.490	&	10.4	\\  
P/2006 U5 (Christensen)	&	17.63	$\pm$	0.01	&	2.04	&	1.538	&	2.358	&	16.6	\\  \hline
\hline
\end{tabular}
\caption[The observed comets]{The observed comets. The $r$ band magnitudes 
reported are SDSS
  model magnitudes except in the case of unresolved comets (denoted by
  a $\dagger$) where the SDSS PSF magnitudes are reported.  $\Delta$, R, and phase list the Geocentric and Heliocentric distance (AU) and phase angle of the comet at the time of observation.}
\label{table_obs}
\end{center}
\end{table}
\normalsize

\begin{table} [p]
\tiny
\begin{tabular}{|l|c|c|c|c|c|}
\hline \hline
Comet	&		$r$		&	$u-g$			&	$g-r$			&	$r-i$			&	$i-z$			\\ \hline
\bf{ACTIVE COMETS}	&				&				&				&				&				\\  
30P Reinmuth	&	14.85	$\pm$	0.01	&	1.57	$\pm$	0.02	&	0.60	$\pm$	0.01	&	0.32	$\pm$	0.01	&	0.19	$\pm$	0.01	\\  
46P/Wirtanen	&	18.84	$\pm$	0.02	&	 	-	 	&	0.54	$\pm$	0.09	&	0.25	$\pm$	0.05	&	0.06	$\pm$	0.05	\\  
47P/Ashbrook-Jackson	&	20.57	$\pm$	0.05	&	1.09	$\pm$	0.33	&	0.59	$\pm$	0.07	&	0.38	$\pm$	0.07	&	0.02	$\pm$	0.17	\\  
50P/Arend	&	18.13	$\pm$	0.03	&	1.72	$\pm$	0.32	&	0.57	$\pm$	0.04	&	0.08	$\pm$	0.04	&	0.27	$\pm$	0.10	\\  
62P/Tsuchinshan	&	15.04	$\pm$	0.01	&	1.27	$\pm$	0.04	&	0.45	$\pm$	0.01	&	0.18	$\pm$	0.01	&	0.07	$\pm$	0.02	\\  
64P/Swift-Gehrels	&	19.23	$\pm$	0.04	&	1.8	$\pm$	0.45	&	0.60	$\pm$	0.06	&	0.22	$\pm$	0.06	&	0.00	$\pm$	0.21	\\  
65P/Gunn (1)	&	17.13	$\pm$	0.01	&	1.68	$\pm$	0.07	&	0.58	$\pm$	0.01	&	0.21	$\pm$	0.01	&	0.08	$\pm$	0.03	\\  
65P/Gunn (2)	&	17.15	$\pm$	0.01	&	1.57	$\pm$	0.06	&	0.55	$\pm$	0.01	&	0.21	$\pm$	0.01	&	0.10	$\pm$	0.03	\\  
67P/Churyumov-Gerasimenko	&	14.28	$\pm$	0.01	&	1.83	$\pm$	0.03	&	0.70	$\pm$	0.01	&	0.25	$\pm$	0.01	&	0.06	$\pm$	0.01	\\  
69P/Taylor	&	15.59	$\pm$	0.01	&	1.51	$\pm$	0.04	&	0.60	$\pm$	0.01	&	0.25	$\pm$	0.01	&	0.06	$\pm$	0.02	\\  
70P/Kojima	&	16.68	$\pm$	0.01	&	1.47	$\pm$	0.07	&	0.64	$\pm$	0.01	&	0.23	$\pm$	0.01	&	0.21	$\pm$	0.03	\\  
129P/Shoemaker-Levy	&	18.07	$\pm$	0.02	&	1.18	$\pm$	0.12	&	0.55	$\pm$	0.03	&	0.17	$\pm$	0.03	&	0.09	$\pm$	0.07	\\  
146P/Shoemaker-Linear	&	18.78	$\pm$	0.02	&	1.46	$\pm$	0.15	&	0.64	$\pm$	0.03	&	0.25	$\pm$	0.03	&	0.12	$\pm$	0.06	\\  
158P/Kowal-LINEAR (1)	&	18.78	$\pm$	0.02	&	1.59	$\pm$	0.19	&	0.56	$\pm$	0.03	&	0.22	$\pm$	0.03	&	0.01	$\pm$	0.09	\\  
158P/Kowal-LINEAR (2)	&	18.60	$\pm$	0.02	&	1.48	$\pm$	0.14	&	0.59	$\pm$	0.03	&	0.19	$\pm$	0.02	&	0.20	$\pm$	0.05	\\  
158P/Kowal-LINEAR (3)	&	19.36	$\pm$	0.02	&	2.09	$\pm$	0.39	&	0.58	$\pm$	0.03	&	0.20	$\pm$	0.04	&	0.19	$\pm$	0.08	\\  
P/2002 EJ57 (LINEAR)	&	18.99	$\pm$	0.01	&	1.57	$\pm$	0.11	&	0.54	$\pm$	0.02	&	0.20	$\pm$	0.02	&	0.12	$\pm$	0.05	\\  
2003 WY25	&	18.33	$\pm$	0.01	&	1.47	$\pm$	0.08	&	0.64	$\pm$	0.01	&	0.27	$\pm$	0.01	&	0.12	$\pm$	0.04	\\  
C/1999 F2 (Dalcanton)	&	15.81	$\pm$	0.01	&	1.69	$\pm$	0.05	&	0.52	$\pm$	0.07	&	0.25	$\pm$	0.05	&	0.07	$\pm$	0.05	\\  
C/2000 K2 (LINEAR)	&	16.82	$\pm$	0.01	&	1.55	$\pm$	0.06	&	0.55	$\pm$	0.01	&	0.22	$\pm$	0.01	&	0.10	$\pm$	0.02	\\  
C/2000 QJ46 (LINEAR) (1)	&	17.44	$\pm$	0.01	&	1.55	$\pm$	0.11	&	0.57	$\pm$	0.02	&	0.23	$\pm$	0.01	&	0.12	$\pm$	0.04	\\  
C/2000 QJ46 (LINEAR) (2)	&	19.39	$\pm$	0.04	&	1.04	$\pm$	0.27	&	0.56	$\pm$	0.06	&	0.41	$\pm$	0.06	&	0.09	$\pm$	0.13	\\  
C/2000 SV74 (LINEAR)	&	14.68	$\pm$	0.00	&	1.66	$\pm$	0.02	&	0.52	$\pm$	0.01	&	0.21	$\pm$	0.01	&	0.04	$\pm$	0.01	\\  
C/2000 Y2 (Skiff)	&	16.71	$\pm$	0.01	&	1.58	$\pm$	0.04	&	0.55	$\pm$	0.01	&	0.18	$\pm$	0.01	&	0.09	$\pm$	0.01	\\  
C/2001 RX14 (LINEAR)	&	12.62	$\pm$	0.01	&	1.61	$\pm$	0.03	&	0.57	$\pm$	0.02	&	0.23	$\pm$	0.02	&	0.06	$\pm$	0.01	\\  
C/2002 O7 (LINEAR) (1)	&	19.37	$\pm$	0.03	&	1.35	$\pm$	0.21	&	0.58	$\pm$	0.04	&	0.14	$\pm$	0.04	&	-0.12	$\pm$	0.25	\\  
C/2002 O7 (LINEAR) (2)	&	15.24	$\pm$	0.01	&	1.56	$\pm$	0.07	&	0.56	$\pm$	0.02	&	0.22	$\pm$	0.02	&	0.07	$\pm$	0.03	\\  
P/1999 V1 (Catalina)	&	17.36	$\pm$	0.01	&	1.73	$\pm$	0.09	&	0.58	$\pm$	0.01	&	0.23	$\pm$	0.01	&	0.09	$\pm$	0.03	\\  
P/2002 T5 (LINEAR)	&	18.33	$\pm$	0.02	&	1.52	$\pm$	0.21	&	0.66	$\pm$	0.03	&	0.23	$\pm$	0.03	&	0.09	$\pm$	0.10	\\  
P/2004 A1 (LONEOS)	&	18.55	$\pm$	0.01	&	1.67	$\pm$	0.20	&	0.61	$\pm$	0.02	&	0.31	$\pm$	0.02	&	0.09	$\pm$	0.05	\\  
P/2006 U5 (Christensen)	&	17.63	$\pm$	0.01	&	1.46	$\pm$	0.09	&	0.51	$\pm$	0.01	&	0.08	$\pm$	0.01	&	0.22	$\pm$	0.04	\\ 
\bf{Median Color}	&				&		$\bf{1.57 \pm 0.21}$		&		$\bf{0.57 \pm 0.05}$		&		$\bf{0.22 \pm 0.07}$		&		$\bf{0.09 \pm 0.07}$		\\ \hline
\bf{UNRESOLVED COMETS}	&		 		&		 		&		 		&		 		&		 		\\  
19P/Borrelly	&	21.35	$\pm$	0.05	&	1.11	$\pm$	0.50	&	0.77	$\pm$	0.09	&	0.12	$\pm$	0.09	&	0.27	$\pm$	0.25	\\  
113P/Spitaler	&	21.05	$\pm$	0.05	&	1.06	$\pm$	0.44	&	0.60	$\pm$	0.08	&	0.33	$\pm$	0.07	&	0.00	$\pm$	0.61	\\  
174P (60558 Echeclus)	&	21.20	$\pm$	0.08	&	1.55	$\pm$	1.02	&	0.73	$\pm$	0.13	&	0.18	$\pm$	0.08	&	-0.71	$\pm$	0.53	\\  
176P  (118401 LINEAR)	&	20.27	$\pm$	0.03	&	1.67	$\pm$	0.40	&	0.54	$\pm$	0.04	&	0.06	$\pm$	0.05	&	0.04	$\pm$	0.10	\\  \hline
\hline
\end{tabular}
\caption[Comet colors]{ The SDSS colors of the comets. The listed $r$ band
  magnitudes are SDSS model magnitudes for active comets, and SDSS PSF
  magnitudes for the unresolved comets. }
\label{table_color}
\end{table}

\begin{table} [p]
\begin{center}
\tiny
\begin{tabular}{|l|c|c|c|c|c|c|}
\hline \hline
Comet	&		Psf-model & \multicolumn{5}{c|}{Surface
  Brightness Profile Slope} \\	& &			$u$			&	$g$			&	$r$			&	$i$			&	$z$			\\  \hline
30P Reinmuth	&		2.20	&			-1.57	$\pm$	0.16	&	-1.57	$\pm$	0.01	&	-1.58	$\pm$	0.01	&	-1.54	$\pm$	0.01	&	-1.55	$\pm$	0.01	\\  
46P/Wirtanen	&		1.63	&			 	-	 	&	-1.18	$\pm$	0.06	&	-1.22	$\pm$	0.06	&	-1.34	$\pm$	0.06	&	-1.54	$\pm$	0.40	\\  
47P/Ashbrook-Jackson	&		0.24	&			 	-	 	&	-1.87	$\pm$	0.26	&	-2.85	$\pm$	0.28	&	-1.74	$\pm$	0.06	&	-2.43	$\pm$	0.31	\\  
50P/Arend	&		2.35	&			 	-	 	&	-1.31	$\pm$	0.07	&	-1.38	$\pm$	0.11	&	-1.50	$\pm$	0.15	&	-1.32	$\pm$	0.29	\\  
62P/Tsuchinshan	&		3.91	&			-0.91	$\pm$	0.07	&	-0.86	$\pm$	0.01	&	-0.97	$\pm$	0.01	&	-0.98	$\pm$	0.01	&	-0.90	$\pm$	0.03	\\  
64P/Swift-Gehrels	&		1.75	&				-		&	-1.29	$\pm$	0.18	&	-1.22	$\pm$	0.11	&	-1.28	$\pm$	0.16	&		-		\\  
65P/Gunn (1)	&		1.75	&			-1.58	$\pm$	0.10	&	-1.88	$\pm$	0.05	&	-1.94	$\pm$	0.04	&	-1.94	$\pm$	0.04	&	-1.66	$\pm$	0.17	\\  
65P/Gunn (2)	&		1.90	&			-2.27	$\pm$	0.56	&	-1.87	$\pm$	0.03	&	-1.91	$\pm$	0.04	&	-1.82	$\pm$	0.04	&	-1.73	$\pm$	0.07	\\  
67P/Churyumov-Gerasimenko	&		3.15	&			-0.99	$\pm$	0.04	&	-1.05	$\pm$	0.01	&	-1.04	$\pm$	0.01	&	-1.02	$\pm$	0.01	&	-1.05	$\pm$	0.02	\\  
69P/Taylor	&		3.19	&			-1.38	$\pm$	0.24	&	-1.24	$\pm$	0.01	&	-1.35	$\pm$	0.01	&	-1.33	$\pm$	0.01	&	-1.42	$\pm$	0.04	\\  
70P/Kojima	&		1.97	&			-0.97	$\pm$	0.41	&	-1.32	$\pm$	0.03	&	-1.41	$\pm$	0.02	&	-1.43	$\pm$	0.02	&	-1.39	$\pm$	0.06	\\  
129P/Shoemaker-Levy	&		2.03	&			 	-	 	&	-1.85	$\pm$	0.10	&	-1.67	$\pm$	0.04	&	-1.53	$\pm$	0.07	&	-1.13	$\pm$	0.20	\\  
146P/Shoemaker-LINEAR	&		1.27	&			 	-	 	&	-1.62	$\pm$	0.12	&	-1.74	$\pm$	0.11	&	-1.74	$\pm$	0.10	&	-1.75	$\pm$	0.42	\\  
158P/Kowal-LINEAR (1)	&		1.15	&			 	-	 	&	-1.99	$\pm$	0.11	&	-2.20	$\pm$	0.10	&	-2.03	$\pm$	0.15	&	-2.10	$\pm$	0.19	\\  
158P/Kowal-LINEAR (2)	&		1.10	&			-2.28	$\pm$	0.59	&	-2.30	$\pm$	0.11	&	-2.30	$\pm$	0.08	&	-2.08	$\pm$	0.10	&	-2.79	$\pm$	0.49	\\  
158P/Kowal-LINEAR (3)	&		1.01	&			 	-	 	&	-2.12	$\pm$	0.17	&	-1.85	$\pm$	0.07	&	-1.83	$\pm$	0.11	&	-1.77	$\pm$	0.33	\\  
P/2002 EJ57 (LINEAR)	&		0.37	&			-1.87	$\pm$	0.36	&	-2.84	$\pm$	0.18	&	-2.96	$\pm$	0.14	&	-3.01	$\pm$	0.12	&	-3.14	$\pm$	0.27	\\  
2003 WY25	&		0.66	&			-1.24	$\pm$	0.39	&	-1.88	$\pm$	0.07	&	-1.76	$\pm$	0.06	&	-2.04	$\pm$	0.09	&	-1.90	$\pm$	0.17	\\  
C/1999 F2 (Dalcanton)	&		2.29	&			-1.12	$\pm$	0.10	&	-1.08	$\pm$	0.01	&	-1.07	$\pm$	0.01	&	-1.03	$\pm$	0.01	&	-1.07	$\pm$	0.03	\\  
C/2000 K2 (LINEAR)	&		1.74	&			-2.03	$\pm$	0.41	&	-1.94	$\pm$	0.03	&	-1.92	$\pm$	0.03	&	-1.89	$\pm$	0.02	&	-1.71	$\pm$	0.05	\\  
C/2000 QJ46 (LINEAR) (1)	&		1.64	&			-2.05	$\pm$	0.35	&	-1.69	$\pm$	0.04	&	-1.73	$\pm$	0.02	&	-1.89	$\pm$	0.04	&	-1.95	$\pm$	0.13	\\  
C/2000 QJ46 (LINEAR) (2)	&		1.47	&			 	-	 	&	-1.96	$\pm$	0.32	&	-1.49	$\pm$	0.21	&	-1.96	$\pm$	0.44	&	-1.96	$\pm$	0.60	\\  
C/2000 SV74 (LINEAR)	&		3.55	&			-1.19	$\pm$	0.17	&	-1.24	$\pm$	0.01	&	-1.23	$\pm$	0.01	&	-1.22	$\pm$	0.01	&	-1.20	$\pm$	0.02	\\  
C/2000 Y2 (Skiff)	&		1.19	&			-2.08	$\pm$	0.22	&	-2.11	$\pm$	0.02	&	-2.17	$\pm$	0.03	&	-2.08	$\pm$	0.04	&	-1.94	$\pm$	0.06	\\  
C/2001 RX14 (LINEAR)	&		2.76	&			-1.31	$\pm$	0.01	&	-1.39	$\pm$	0.00	&	-1.40	$\pm$	0.01	&	-1.36	$\pm$	0.00	&	-1.35	$\pm$	0.01	\\  
C/2002 O7 (LINEAR) (1)	&		1.23	&			 	-	 	&	-2.41	$\pm$	0.10	&	-2.49	$\pm$	0.28	&	-2.01	$\pm$	0.22	&	-1.06	$\pm$	0.34	\\  
C/2002 O7 (LINEAR) (2)	&		2.94	&			-1.37	$\pm$	0.05	&	-1.49	$\pm$	0.04	&	-1.47	$\pm$	0.04	&	-1.40	$\pm$	0.05	&	-1.37	$\pm$	0.05	\\  
P/1999 V1 (Catalina)	&		1.87	&			-1.03	$\pm$	0.06	&	-1.64	$\pm$	0.03	&	-1.60	$\pm$	0.03	&	-1.65	$\pm$	0.03	&	-1.55	$\pm$	0.10	\\  
P/2002 T5 (LINEAR)	&		1.97	&			 	-	 	&	-1.68	$\pm$	0.11	&	-1.70	$\pm$	0.08	&	-1.51	$\pm$	0.05	&	-1.34	$\pm$	0.53	\\  
P/2004 A1 (LONEOS)	&		0.99	&			-	&	-2.41	$\pm$	0.09	&	-2.51	$\pm$	0.13	&	-2.44	$\pm$	0.14	&	-1.98	$\pm$	0.29	\\  
P/2006 U5 (Christensen)	&		2.04	&			-1.87	$\pm$	0.48	&	-1.57	$\pm$	0.05	&	-1.59	$\pm$	0.03	&	-1.70	$\pm$	0.03	&	-1.46	$\pm$	0.06	\\  \hline

\hline
\end{tabular}
\caption[Surface brightness profile fits]{Linear fits of the Surface Brightness Profile slopes of the active comets in all five SDSS bands, described in Section \ref{sec_SBP}. PSF-model is the difference between the SDSS measured PSF and model magnitudes}
\label{table_slope}
\end{center}
\end{table}

\begin{table} [p]
\begin{center}
\tiny
\begin{tabular}{|l|c|c|c|c|c|c|}
\hline \hline															
Comet	&	QC	&	A$(\alpha)f\rho$			&	A$f(0)\rho$			&	SDSS radius (km)	&	Radius (km)	&	Reference	\\ \hline
$\dagger$19P/Borrelly (JFC)	&	-	&		-		&		-	 	&	2.99	&	1.8 $-$ 4.4	&	[1]	\\
30P Reinmuth	&	1	&	57	$\pm$	2.4	&	98.5	$\pm$	4.1	&	5.31	&	1	&	[2]	\\
46P/Wirtiran	&	1	&	9.6	$\pm$	2.3	&	11.4	$\pm$	2.8	&	1.17	&	0.58	&	[2]	\\
47P/Ashbrook-Jackson	&	2	&	26	$\pm$	2.3	&	28.3	$\pm$	2.5	&	3.56	&	3.38	&	[3]	\\
50P/Arend	&	1	&	12.5	$\pm$	2.5	&	18.5	$\pm$	3.6	&	2.44	&	0.95	&	[4]	\\
62P/Tsuchinshan	&	1	&	17.5	$\pm$	3.1	&	22.7	$\pm$	4.1	&	1.09	&	-	&	[2]	\\
64P/Swift-Gehrels	&	1	&	10.3	$\pm$	1.2	&	14.8	$\pm$	1.8	&	1.9	&	1.83	&	[2]	\\
65P/Gunn (1)	&	1	&	212	$\pm$	22.9	&	256	$\pm$	27.6	&	8.93	&	4.59	&	[2]	\\
65P/Gunn (2)	&	1	&	189.3	$\pm$	30.4	&	229.4	$\pm$	36.8	&	8.3	&	4.59	&	[2]	\\
67P/Churyumov-Gerasimenko	&	1	&	67.2	$\pm$	2.9	&	118.4	$\pm$	5.1	&	4.84	&	2.39	&	[5]	\\
69P/Taylor	&	1	&	21	$\pm$	4.6	&	31.1	$\pm$	6.8	&	1.63	&	2.1	&	[2]	\\
70P/Kojima	&	1	&	29.1	$\pm$	3.5	&	34.5	$\pm$	4.2	&	2.67	&	2.18	&	[6]	\\
$\dagger$113P/Spitaler (JFC)	&	-	&		-		&		-		&	1.49	&	1.15	&	[2]	\\
129P/Shoemaker-Levy	&	1	&	21.3	$\pm$	2.7	&	29.5	$\pm$	3.7	&	2.87	&	1.66	&	[2]	\\
146P/Shoemaker-Linear	&	2	&	4.4	$\pm$	0.3	&	6.8	$\pm$	0.4	&	1.14	&	-	&		\\
158P/Kowal-LINEAR (1)	&	2	&	78.3	$\pm$	9.2	&	85.9	$\pm$	10.1	&	5.42	&	-	&		\\
158P/Kowal-LINEAR (2)	&	2	&	98.6	$\pm$	25.3	&	110.4	$\pm$	28.3	&	6.2	&	-	&		\\
158P/Kowal-LINEAR (3)	&	2	&	43	$\pm$	6.6	&	50.7	$\pm$	7.8	&	5.46	&	-	&		\\
$\dagger$174P (60558 Echeclus)	&	-	&		-		&		-		&	39.25	&	41.8	&	[7]	\\
$\dagger$176P (118401 LINEAR)	&	-	&		-		&		-		&	2.36	&	-	&		\\
P/2002 EJ57 (LINEAR)	&	2	&	24.9	$\pm$	3.8	&	28.6	$\pm$	4.4	&	2.12	&	-	&		\\
2005 WY25	&	2	&	0.9	$\pm$	0	&	1.4	$\pm$	0.1	&	0.23	&	0.16	&	[8]	\\
C/1999 F2 (Dalcanton)	&	1	&	817.7	$\pm$	146.7	&	997.6	$\pm$	179	&	18.18	&	-	&		\\
C/2000 K2 (LINEAR)	&	1	&	421.6	$\pm$	71.5	&	496.5	$\pm$	84.2	&	12.36	&	-	&		\\
C/2000 QJ46 (LINEAR) (1)	&	1	&	11.4	$\pm$	0.5	&	12.4	$\pm$	0.5	&	1.2	&	-	&		\\
C/2000 QJ46 (LINEAR) (2)	&	2	&	23.2	$\pm$	1.2	&	31.2	$\pm$	1.6	&	3.14	&	-	&		\\
C/2000 SV74 (LINEAR)	&	1	&	553.1	$\pm$	111.5	&	713.2	$\pm$	143.8	&	14.63	&	-	&		\\
C/2000 Y2 (Skiff)	&	2	&	95.6	$\pm$	10.4	&	109.9	$\pm$	11.9	&	4.41	&	-	&		\\
C/2001 RX14 (LINEAR)	&	1	&	520.3	$\pm$	53.3	&	844.8	$\pm$	86.5	&	13.25	&	-	&		\\
C/2002 O7 (LINEAR) (1)	&	2	&	152.5	$\pm$	35.8	&	187.7	$\pm$	44	&	8.87	&	-	&		\\
C/2002 O7 (LINEAR) (2)	&	1	&	200.4	$\pm$	18.6	&	279.9	$\pm$	25.9	&	7	&	-	&		\\
C/2002 T5 (LINEAR)	&	1	&	92.3	$\pm$	14.5	&	112.2	$\pm$	17.7	&	6.6	&	-	&		\\
P/1999 V1 (Catalina)	&	1	&	40.9	$\pm$	3.9	&	48.3	$\pm$	4.7	&	3.23	&	-	&		\\
P/2004 A1 (LONEOS)	&	2	&	255.4	$\pm$	20.1	&	317.4	$\pm$	25	&	12.6	&	-	&		\\
P/2006 U5 (Christensen)	&	1	&	15.3	$\pm$	1.2	&	21	$\pm$	1.7	&	1.66	&	-	&		\\ \hline
\hline
\end{tabular}
\caption[Radii and A$f\rho$]{Derived quantities. A$f\rho$ values are
  those derived
  from SDSS observations; A$(0)f\rho$ entries are corrected for a
  phase angle of $\alpha=0$ \citep[see][]{1981ESASP.174...47D,
  2010arXiv1001.3010A}. The SDSS radii are all upper limits except for the
  unresolved comets marked with a $\dagger$.  
  Literature radii values from 
  [1]	\cite{1998A&A...337..945L}, [2] \cite{2006Icar..182..527T}, 
  [3] \cite{2006MNRAS.373.1590S}, [4] \cite{2009A&A...508.1045L}, 
  [5] \cite{2008A&A...490..377T}, [6] \cite{2008MNRAS.385..737S}, 
  [7] \cite{2008ssbn.book..161S}, and [8] \cite{2006AJ....131.2327J} 
  }
\label{table_derived}
\end{center}
\end{table}

\begin{table} [p]
\begin{center}
\scriptsize
\begin{tabular}{|l|c|c|c|c|c|c|}
\hline \hline
Population	&	$u-g$	&	$g-r$	&	$r-i$	&	$i-z$	&	$S^{'}$ ($u,g,r,i,z$)	&	$S^{'}$ ($g,r,i$)	\\ \hline
Comets	&	$1.57 \pm 0.21$	&	$0.57 \pm 0.05$	&	$0.22 \pm 0.07$	&	$0.09 \pm 0.07$	&	8.4 $\pm$ 3.6	&	8.3$\pm$3.5	\\ 
Trojans	&	$1.47 \pm 0.17$	&	$0.58 \pm 0.07$	&	$0.23 \pm 0.5$	&	$0.13 \pm 0.08$	&	$8.2 \pm 2.9$	&	8.8$\pm$2.9	\\ 
Centaurs	&	1.41 $\pm$ 0.46	&	0.62$\pm$0.16	&	0.26$\pm$0.17	&	0.21$\pm$0.16	&	8.9$\pm$7.7	&	10.3$\pm$11.1	\\ 
TNO	&	$1.03 \pm 1.21$ 	&	$0.74 \pm 0.40$	&
$0.35 \pm 0.26$	& $0.18 \pm 0.17$	&	$16.4 \pm 9.2$	&
$17.8 \pm 11.8$	\\ \hline
\hline
\end{tabular}
\caption[Colors and spectral gradient of small body
  populations]{Colors and spectral gradient of populations as measured
  by the SDSS. The colors of the comets, centaurs, and Trojans, and TNOs here
  are those of the bodies used to calculate the $S^{'}$, as discussed
  in section \ref{sec_dis_color}. The TNO colors are from \cite{becker}. }
\label{table_sprime}
\end{center}
\end{table}

\begin{table}[b!]
\begin{center}
\scriptsize
\begin{tabular}{|l|c|c|}
\hline \hline
Reference	&	JFC CLF	 		&	JFC CSD	 		\\ \hline
This Work	&	0.49	$\pm$	0.05	&	-2.4	$\pm$	0.2	\\ 
\cite{2006Icar..182..527T}	&	0.53	$\pm$	0.05	&	-2.7	$\pm$	0.3	\\  
\cite{2006LPICo1325...76W}	&	0.35	$\pm$	0.01	&	-1.73	$\pm$	0.06	\\ 
\cite{2004come.book..223L}	&	0.38	$\pm$	0.06	&	-1.9	$\pm$	0.3	\\  
\cite{LIII}	&	0.32	$\pm$	0.02	&	-1.6	$\pm$	0.1	\\  
\cite{Fer}	&	0.53	$\pm$	0.05	&	-2.65	$\pm$	0.25	\\ \hline
\hline
\end{tabular}
\caption[Cumulative luminosity function and size distribution for JFC]{Estimates of the cumulative luminosity function and
  cumulative size distribution power law slopes for the Jupiter Family Comets.
}
\label{table_CLF}
\end{center}
\end{table}

\begin{table} [p]
\begin{center}
\scriptsize
\begin{tabular}{|l|c|c|c|l|}
\hline \hline
Population	&	Broken Min	&	Single Power Law	&
Broken Max	&	Source	\\ \hline
$\dagger$Jupiter Family Comets	&	0.19 $\pm$ 0.03	&	0.49 $\pm$ 0.05	&	0.73 $\pm$ 0.08	&	This Work	\\ 
Near Earth Asteroids	&		&	$0.35 \pm 0.02$	&		&	\cite{2000Sci...288.2190B}	\\ 
Asteroids (cometary orbits)	&		&	0.51 $\pm$ 0.02	&		&	\cite{CLF}	\\ 
$\dagger$Main Belt Asteroids (broken)	&	0.10-0.59	&	0.35-0.97	&	0.37-1.04	&	\cite{2008Icar..198..138P}	\\  
Hildas	&		&	0.42 $\pm$ 0.02	&		&	\cite{CLF}	\\  
$\dagger$Trojans	&		&	0.44 $\pm$ 0.05	&		&	\cite{2007MNRAS.377.1393S}	\\  
Centaurs	&		&	0.54 $\pm$ 0.07	&		&	\cite{2001AJ....121..562L}	\\ 
Kuiper Belt Objects	&		&	$0.66 \pm 0.06$	&		&	\cite{2001AJ....122.2740T}	\\ \hline

\hline
\end{tabular}
\caption[Cumulative luminosity function for small-body populations]{Estimates of the cumulative luminosity function power
  laws for small-body populations.
  The CLF of Main Belt
  Asteroid are family populations that are fit both by broken and unbroken
  power laws, with the values cited here being the ranges determined for
  these families in \cite{2008Icar..198..138P}.  Those
  populations marked by a $\dagger$ are derived from SDSS data.}
\label{table_popCLF}
\end{center}
\end{table}

\begin{table} [p]
\scriptsize
\begin{tabular}{|l|c|c|c|c|c|}
\hline \hline
Comet	&	R (AU)	&	A$f\rho$ (cm)	&	Previous R
(AU)	&	Previous A$f\rho$ (cm)	&	Source	\\ \hline
46P/Wirtanen	&	2.598	&	9.59 $\pm$ 2.35	&	1.120	&	112.2	&	1	\\  
47P/Ashbrook-Jackson	&	4.572	&	25.97 $\pm$ 2.28	&	4.030	&	28.95 $\pm$ 4.1	&	2	\\  
65P/Gunn	&	4.340	&	211.97 $\pm$ 22.86	&	4.430	&	133.4 $\pm$ 4.7	&	3	\\  
	&	4.338	&	189.28 $\pm$ 30.37	&	2.640	&	23.4	&	1	\\ 
$\dagger$67P/Churyumov-Gerasimenko	&	1.836	&	67.21 $\pm$ 2.9	&	1.380	&	208.9	&	1	\\  
69P/Taylor	&	1.950	&	21 $\pm$ 4.59	&	4.030	&	18.39 $\pm$ 1.7	&	2	\\  
129P/Shoemaker-Levy	&	3.051	&	21.26 $\pm$ 2.68	&	4.558	&	4.5 $\pm$ 0.6	&	4	\\ \hline
\hline
\end{tabular}
\caption[A$f\rho$ in literature]{A comparison of our A$f\rho$ values with previous listed
  values. References: [1] \cite{1995Icar..118..223A}, [2]
  \cite{LIII}, [3] \cite{2001A&A...365..204L}, [4]
  \cite{2005MNRAS.358..641L}. ($\dagger$A detailed look at
  A$f\rho$ for 67P/Churyumov-Gerasimenko can be found in
  \cite{2010arXiv1001.3010A}.} 
\label{table_Afp}
\label{lasttable}
\end{table}

\normalsize
}

\clearpage


\clearpage

\end{document}